\documentclass[reprint,prl,aps]{revtex4-1}
\usepackage{graphicx}
\usepackage{color}
\usepackage{dcolumn}
\usepackage{bm}
\usepackage{amssymb}
\usepackage{color}
\usepackage{soul,xcolor} 
\setstcolor{red} 
\usepackage{epstopdf}

\newcolumntype{L}[1]{>{\raggedright\arraybackslash}m{#1}}
\newcolumntype{C}[1]{>{\centering\arraybackslash}m{#1}}
\newcolumntype{R}[1]{>{\raggedleft\arraybackslash}m{#1}}
\newcolumntype{N}{@{}m{0pt}@{}}

\begin{document}

\title{Tuning superconductivity in twisted bilayer graphene}

\author{Matthew Yankowitz$^{1*}$}
\author{Shaowen Chen$^{1,2*}$}
\author{Hryhoriy Polshyn$^{3*}$} 
\author{K. Watanabe$^{4}$} 
\author{T. Taniguchi$^{4}$} 
\author{David Graf$^{5}$}
\author{Andrea F. Young$^{3\dagger}$} 
\author{Cory R. Dean$^{1\dagger}$}

\affiliation{$^{1}$Department of Physics, Columbia University, New York, NY, USA}
\affiliation{$^{2}$Department of Applied Physics and Applied Mathematics, Columbia University, New York, NY, USA}
\affiliation{$^{3}$Department of Physics, University of California, Santa Barbara, CA 93106}
\affiliation{$^{4}$National Institute for Materials Science, 1-1 Namiki, Tsukuba 305-0044, Japan}
\affiliation{$^{5}$National High Magnetic Field Laboratory, Tallahassee, FL 32310}
\affiliation{$^{*}$These authors contributed equally to this work.}
\affiliation{$^{\dagger}$ afy2003@ucsb.edu (A.F.Y.); cd2478@columbia.edu (C.R.D.)}

\date{\today}

\begin{abstract}
Materials with flat electronic bands often exhibit exotic quantum phenomena owing to strong correlations. Remarkably, an isolated low-energy flat band can be induced in bilayer graphene by simply rotating the layers to 1.1$^{\circ}$, resulting in the appearance of gate-tunable superconducting and correlated insulating phases. Here, we demonstrate that in addition to the twist angle, the interlayer coupling can also be modified to precisely tune these phases. We establish the capability to induce superconductivity at a twist angle larger than 1.1$^{\circ}$ --- in which correlated phases are otherwise absent --- by varying the interlayer spacing with hydrostatic pressure. Realizing devices with low disorder additionally reveals new details about the superconducting phase diagram and its relationship to the nearby insulator. Our results demonstrate twisted bilayer graphene to be a uniquely tunable platform for exploring novel correlated states.
\end{abstract}

\maketitle

\begin{figure*}[ht]
\includegraphics[width=6.9 in]{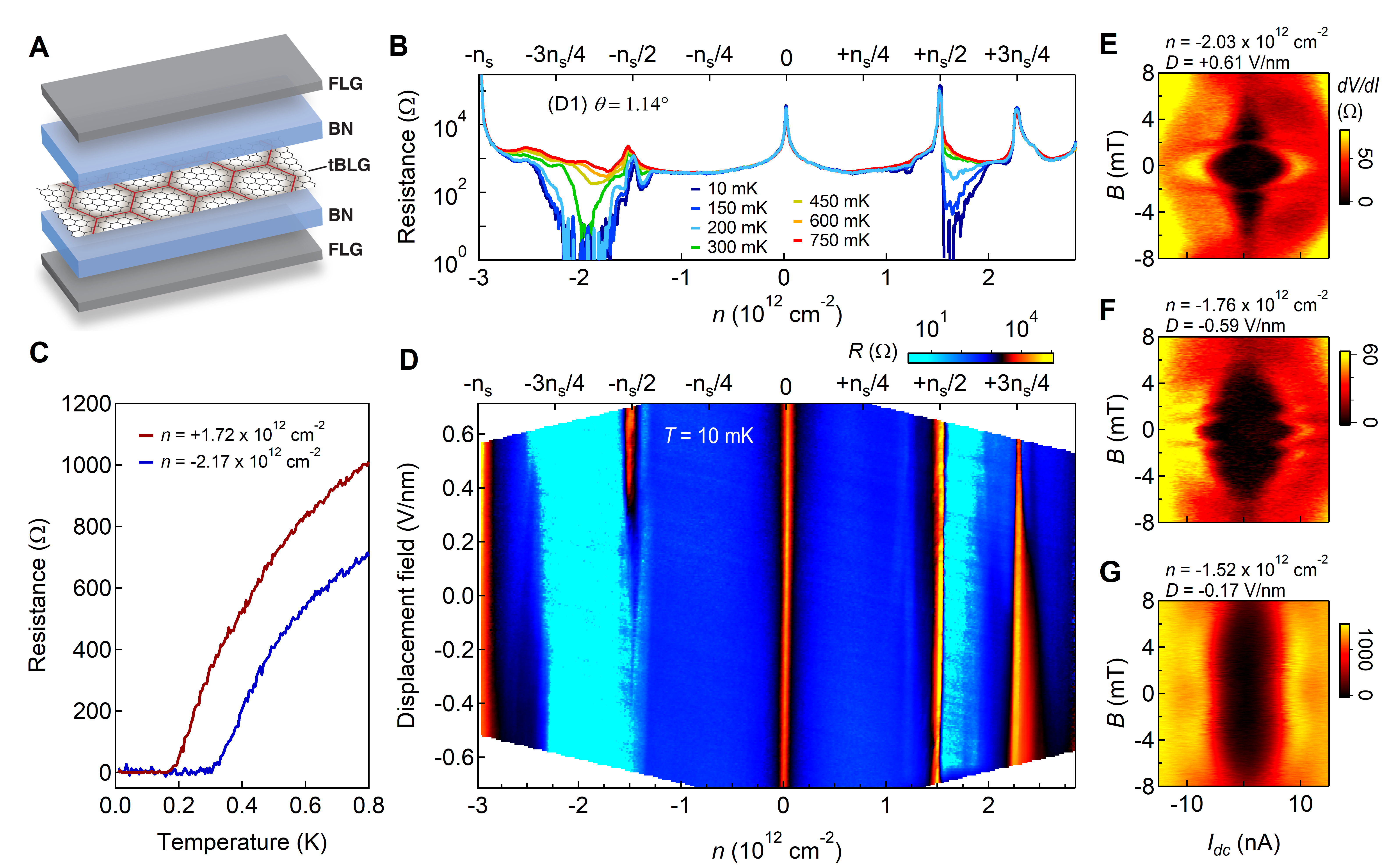} 
\caption{\textbf{Superconductivity in a 1.14$^{\circ}$ device.}
(\textbf{A}) Schematic of an all-van der Waals tBLG heterostructure. tBLG is encapsulated between flakes of BN, with encapsulating flakes of few-layer graphite (FLG) acting as gates. 
(\textbf{B}) Temperature dependence of the resistance of device D1 over the density range necessary to fill the moir\'e unit cell, $n\in[-n_s,n_s]$ at $D$ = 0. The resistance drops to zero over a finite range of $n$ for electron ($n > n_s/2$) and hole ($n < -n_s/2$) doping. 
(\textbf{C}) Resistance as a function of temperature at optimal doping of the hole- and electron-doped superconductors in blue and red, respectively.
(\textbf{D}) Resistance of device D1 as a function of displacement field. At $-n_s/2$ an insulating phase develops at positive $D$, while a superconducting phase develops at negative $D$.
(\textbf{E-G}) Fraunhofer-like quantum interferences of the critical current, arising from one or more Josephson weak links within the sample, measured at: 
(E) $n = -2.03 \times 10^{12}$ cm$^{-2}$ and $D = +0.61$ V/nm,
(F) $n = -1.76 \times 10^{12}$ cm$^{-2}$ and $D = -0.59$ V/nm and
(G) $n = -1.52 \times 10^{12}$ cm$^{-2}$ and $D = -0.17$ V/nm. 
A $\pi$-junction is observed in (G). The measured temperature is $T \approx$ 10 mK for all data sets unless otherwise noted.
}
\label{fig:1}
\end{figure*}

The electronic properties of many materials are well described by assuming that non-interacting electrons simply fill the available energy bands. However, for systems with narrowly dispersing flat bands in which the kinetic energy is small relative to the Coulomb energy, the assumption that electrons are non-interacting is no longer valid. Instead, the electronic ground state is driven by minimizing the mutual Coulomb repulsion between electrons.  This can lead to the emergence of correlated electron phases, often with interesting new properties. It has recently been demonstrated that heterostructures consisting of layered two-dimensional materials provide a remarkably simple avenue to reach this condition, where narrow isolated bands can be realized simply by tuning the rotational ordering between layers~\cite{Bistritzer2011,Chen2018,Wu2018a,Zhang2018}. Bilayer graphene, which normally consists of two vertically stacked monolayer graphene layers arranged in an AB (Bernal) stacking configuration, provides a dramatic example. Upon rotating the layers away from Bernal stacking to the so-called ``magic angle'' of $\sim1.1^{\circ}$, the interplay between the resulting moir\'e superlattice and hybridization between the layers leads to the formation of an isolated flat band at the charge neutrality point~\cite{Bistritzer2011}. Near this flat band angle, recent experiments have observed correlated insulator phases at half band filling~\cite{Cao2018a} and superconductivity upon doping slightly away from half band filling for hole-type carriers~\cite{Cao2018b}.

The discovery of superconductivity in twisted bilayer graphene (tBLG) has sparked intense interest owing in part to the possibility that it arises from an unconventional electron-mediated pairing mechanism (see SI for an overview of the recent theoretical literature). The material composition is remarkably simple, comprising only carbon atoms. Unlike most unconventional superconductors, where exploring different carrier density requires growing different samples, in tBLG the entire correlated phase diagram can be accessed in a single device by field effect gating. Additionally, the available degrees of freedom in tBLG, including twist angle control~\cite{Bistritzer2011}, interlayer separation~\cite{Carr2018,Chittari2018}, and displacement field-induced layer imbalance~\cite{Bistritzer2011,Laissardiere2016}, provide opportunities to experimentally tune the electronic structure in ways that are difficult or impossible to access in previously investigated superconductors. 

Here we present measurements of both the superconducting and correlated insulating states in tBLG at the flat-band condition. We study three separate devices, fabricated in a fully-encapsulated, dual-graphite gate structure in an attempt to enhance device mobility and minimize effects of charge inhomogeneity~\cite{Zibrov2017}. In two devices, which have twist angles close to 1.1$^{\circ}$, we observe new correlated phases, including a superconducting pocket near half-filling of the electron-doped band (previously superconductivity was only observed in the hole-doped band) and resistive states at quarter-filling of both bands that emerge in a magnetic field. Varying the layer imbalance with displacement field reveals an unexpected competition between superconductivity and the correlated insulator phases, which we interpret to arise from differences in disorder of each graphene layer. In-plane magnetic-field dependence suggests a complex interplay between spin and valley degrees of freedom. 

The third device is fabricated at an angle of 1.27$^{\circ}$, which is sufficiently far from the flat-band angle such that signatures of correlated states are largely absent. However, we demonstrate the ability to recover the flat-band condition and observe strongly correlated phases at this angle by applying hydrostatic pressure to reduce the separation between the two graphene layers~\cite{Yankowitz2018}. In the superconducting phase, the critical temperature $T_c$ increases above 3 K at $\sim$1.3 GPa and then diminishes with further pressure. The optimal pressure observed at this angle is in excellent agreement with theoretical efforts to model the relationship between bandwidth and interlayer spacing~\cite{Carr2018,Chittari2018}, establishing layer compression as a viable route to engineer the bandwidth in this system. Moreover, this device shows much higher homogeneity, revealing new details about the phase diagram of the superconducting and correlated insulating states. High resolution magnetoresistance oscillations in this device lend new insights into the structure of the flat electronic band and its relationship to the observed correlated states.

\bigskip
\noindent\textbf{Dual-gated twisted bilayer graphene}\\
Fig. 1A shows a schematic of our device structure. We fabricate BN-encapsulated tBLG using the ``tear and stack'' method to control the graphene alignment~\cite{Kim2016}, and additionally include top and bottom graphite gates. The all-van der Waals device geometry has previously been shown to significantly increase charge homogeneity as compared with the evaporated metal or degenerately doped silicon gates~\cite{Zibrov2017}. Additionally, the dual-gate structure allows us to independently vary the charge carrier density and transverse displacement field, and experimentally investigate the effects of interlayer bias on the correlated states~\cite{Bistritzer2011,Laissardiere2016,Kim2017}.

Fig.~\ref{fig:1}B shows the density-dependent resistance of device D1, with interlayer twist angle $\theta \approx$ 1.14$^{\circ}$. The device exhibits low charge carrier inhomogeneity of $\delta n < 2 \times 10^{10}$ cm$^{-2}$ measured by the full width at half maximum of the resistance peak at the charge neutrality point (CNP). The low charge disorder is further confirmed by the emergence of fractional quantum Hall states at magnetic fields as low as $\sim$4 T~\cite{SI}. In Fig.~\ref{fig:1}B, the resistance is plotted over nearly the full density range of the flat band. We identify the boundaries of the flat band by the appearance of strongly insulating response at densities symmetrically located around the CNP~\cite{SI}. We then use a normalized density scale to define partial band filling with $+n_{s}$ and $-n_{s}$ corresponding the electron- and hole-doped band edges with $\pm 4$ electrons per moir\'e unit cell, respectively. Within the flat band we observe resistive states at the CNP, as well as at $\pm n_s/2$ and $+3n_s/4$. At base temperature of $\sim10$ mK, regions of superconductivity appear both in the hole- and electron-doped regions, with the resistance dropping to zero for densities near $\pm n_s/2$.

In the hole band, the density and magnetic field dependence of the superconducting response  resembles previous observations~\cite{Cao2018b}, with two low resistance pockets flanking the correlated insulating state at $-n_{s}/2$ at intermediate temperatures but with the insulating response growing weaker at base temperature~\cite{SI}. In both bands superconductivity appears more robust on the high density side of the insulator, $|n|>|\pm n_s/2|$, and is much weaker or absent on the low density side, $|n|<|\pm n_s/2|$. In the hole band the low density pocket is not fully superconducting at 10 mK, while in the electron band no signature of superconductivity appears at all down to base temperature, for similar doping range. 

Fig.~\ref{fig:1}C shows the resistance versus temperature measured at optimal doping of both the hole- and electron-type superconducting pockets. The critical temperature is $\sim$0.25 K for electron-type carriers and $\sim$0.4 K for hole-type carriers, defined as the crossover point of linear fits to the low and high temperature portions of the resistance curves on a logarithmic scale. The hole-band $T_c$ is similar to that reported previously for a similar twist angle (Ref.~\onlinecite{Cao2018b}), however this is the first observation of superconductivity for electron-type carriers in tBLG. While the band structure of tBLG is not anticipated to be precisely particle-hole symmetric, the observation of superconductivity over similar ranges of density for both electron and hole carriers suggests a likely connection between the mechanism driving the superconducting phases of both carrier types.

The transport properties of graphene bilayers are generically strongly tunable with $D$~\cite{Zhang2009}, however theoretical modeling of flat-band tBLG suggests that transverse field should have little to no consequence due to the strong interlayer hybridization~\cite{Bistritzer2011,Laissardiere2016,Kim2017}. Fig.~\ref{fig:1}D shows the device resistance as a function of displacement field. The superconducting regions appear to be largely insensitive to $D$, exhibiting for instance similar $T_c$ at very positive and negative values of $D$~\cite{SI}. In contrast, the insulating state at $-n_s/2$ shows a strong dependence on $D$, with resistance exceeding 10~k$\Omega$ for positive $D$ but appearing to drop to zero for negative $D$. Similarly, the peak resistance at $+3n_s/4$ in the electron band also varies strongly but exhibits the opposite dependence, becoming less resistive at large, positive $D$. This opposite trend with $D$ between the electron and hole bands suggests that the insulating state is suppressed when the carriers of either sign are polarized towards the same (top) graphene layer. This response is unexpected and likely to be extrinsic, as freestanding tBLG is symmetric under layer interchange, which corresponds to inversion of the sign of $D$.

\bigskip
\noindent\textbf{Influence of disorder}\\
We conjecture that the superconductor to insulator transition observed at $-n_s/2$ is a consequence of disorder. In particular, polarizing carriers to a more disordered graphene layer may favor formation of a percolating superconducting network that short circuits the insulating phase at $-n_s/2$. While this may arise as a consequence of charge disorder, the measured charge fluctuations are  $\sim 10^{10}$ cm$^{-2}$, insufficient to mix these phases across their native separation in carrier density, which is an order of magnitude larger. We do, however, observe transport signatures of spatial inhomogeneity in the period of the moir\'e pattern, marked by significant variations in the densities of the full-filling insulating states at $\pm n_s$ measured between different pairs of contacts in the sample~\cite{SI}. This is consistent with recent TEM imaging of similar tBLG devices~\cite{Yoo2018}, suggesting spatial variations in the moir\'e period may be ubiquitous in these structures. Moreover, we note that the top and bottom BN layers are randomly oriented --- different rotational alignments to the top and bottom graphene layers, as well as inhomogeneity in the structural relaxation for a given graphene/hBN interface, could result in a layer-asymmetric response under applied $D$.  

\begin{figure*}[ht]
\includegraphics[width=4.6 in]{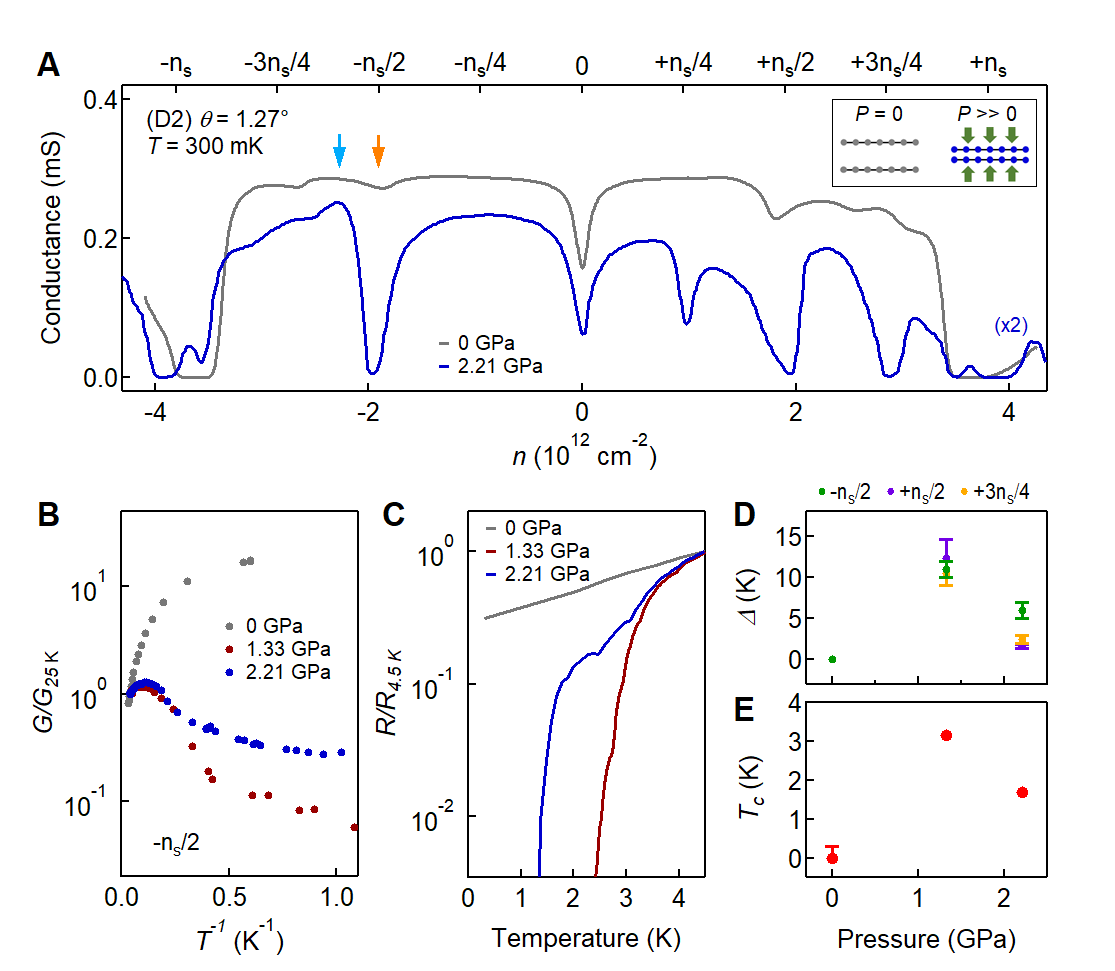} 
\caption{\textbf{Driving superconductivity and correlated insulating states with pressure.}
(\textbf{A}) Conductance of device D2 (1.27$^{\circ}$) measured over the entire density range necessary to fill the moir\'e unit cell at 0 GPa (gray) and 2.21 GPa (blue) at $T$ = 300 mK. Correlated insulating phases are only very weakly resistive at 0 GPa, but develop into strongly insulating states at high pressure. The conductance is measured in a two-terminal voltage bias configuration and includes the contact resistance. (Inset) Cartoon schematic illustrating the decrease in interlayer spacing of the tBLG under high pressure.
(\textbf{B}) Device conductance versus $T^{-1}$ at $-n_s/2$, normalized to its value at $T$ = 25 K (acquired at the density of the orange arrow in panel A).
(\textbf{C}) Four-terminal device resistance versus $T$ for hole doping slightly larger than $-n_s/2$, normalized to its value at $T$ = 4.5 K. The device is a metal at 0 GPa, but becomes a superconductor at high pressure. The two curves at high pressure are taken at optimal doping of the superconductor, and the curve at 0 GPa is taken at the same density as the 1.33 GPa curve (i.e. acquired roughly at the density of the blue arrow in panel A).
(\textbf{D}) Energy gaps $\Delta$ of the correlated insulating phases versus pressure, extracted from the thermal activation measurements in (B) and fit according to $G(T) \propto e^{-\frac{\Delta}{2 k T}}$, where $k$ is the Boltzmann constant. Error bars in the gaps represent the uncertainty arising from determining the linear (thermally activated) regime for the fit.
(\textbf{E}) $T_c$ of the superconducting phase versus pressure. $T_c$ is defined as the crossover point between low and high temperature linear fits to the curves in (C). The upper bound for the 0 GPa curve represents the base temperature of the fridge.
}
\label{fig:2}
\end{figure*}

Evidence of a percolating superconducting network is provided by measurements of the differential resistance $dV/dI$ as a function of applied current $I_{dc}$ and $B$. Fig.~\ref{fig:1}E-G shows three different measurements, sampling different regions of the $D$ versus $n$ response shown in Fig.~\ref{fig:1}D. Periodic oscillations in the critical current $I_c$, resembling Fraunhofer interference, suggest quantum phase coherent transport arising from interspersed regions of superconducting and metallic/insulating phases within the device. The period of the oscillations $\Delta B$ varies from 2 to 4 mT, indicating an effective junction area of $S \approx$ 0.5-1 $\mu$m$^2$, using $S = \Phi_0/\Delta B$, where $\Phi_0 = h/2e$ is the superconducting  flux quantum, $h$ is Planck's constant, and $e$ is the charge of the electron. This constitutes as much as $\sim$40\% of the device area. 

The variations among the quantum interference patterns appearing in Figs.~\ref{fig:1}E-G confirm that the microscopic structure of superconducting regions are tunable with $n$ and $D$. Strikingly, Fig.~\ref{fig:1}G, measured at $-n_s/2$ and negative $D$, shows a minimum in $I_c$ at $B$ = 0, with $I_c$ increasing to a maximum near $\pm4$ mT, indicative of a Josephson junction with $\pi$ phase in its ground state. $\pi$-junctions most commonly arise in two situations: in Josephson junctions consisting of conventional superconductors with a ferromagnetic weak link~\cite{Ryazanov2001}, and in superconductors with sign-changing order parameters~\cite{Tsuei2000}. Either scenario may apply to our data.  The first scenario is consistent with the  observation  that the $\pi$-junction is observed at $-n_s/2$, where non-superconducting regions of the sample are likely in a correlated, presumably magnetic insulator which may break pairs in the superconductor. The second scenario is consistent with theoretical predictions of unconventional pairing symmetry, including $d$-wave pairing. Our results suggest that gate-tuned Josephson junctions realized in more homogeneous devices may prove a powerful tool for investigating the nature of the superconductor and its relationship to the correlated insulator. 

\begin{figure*}[ht]
\includegraphics[width=4.6 in]{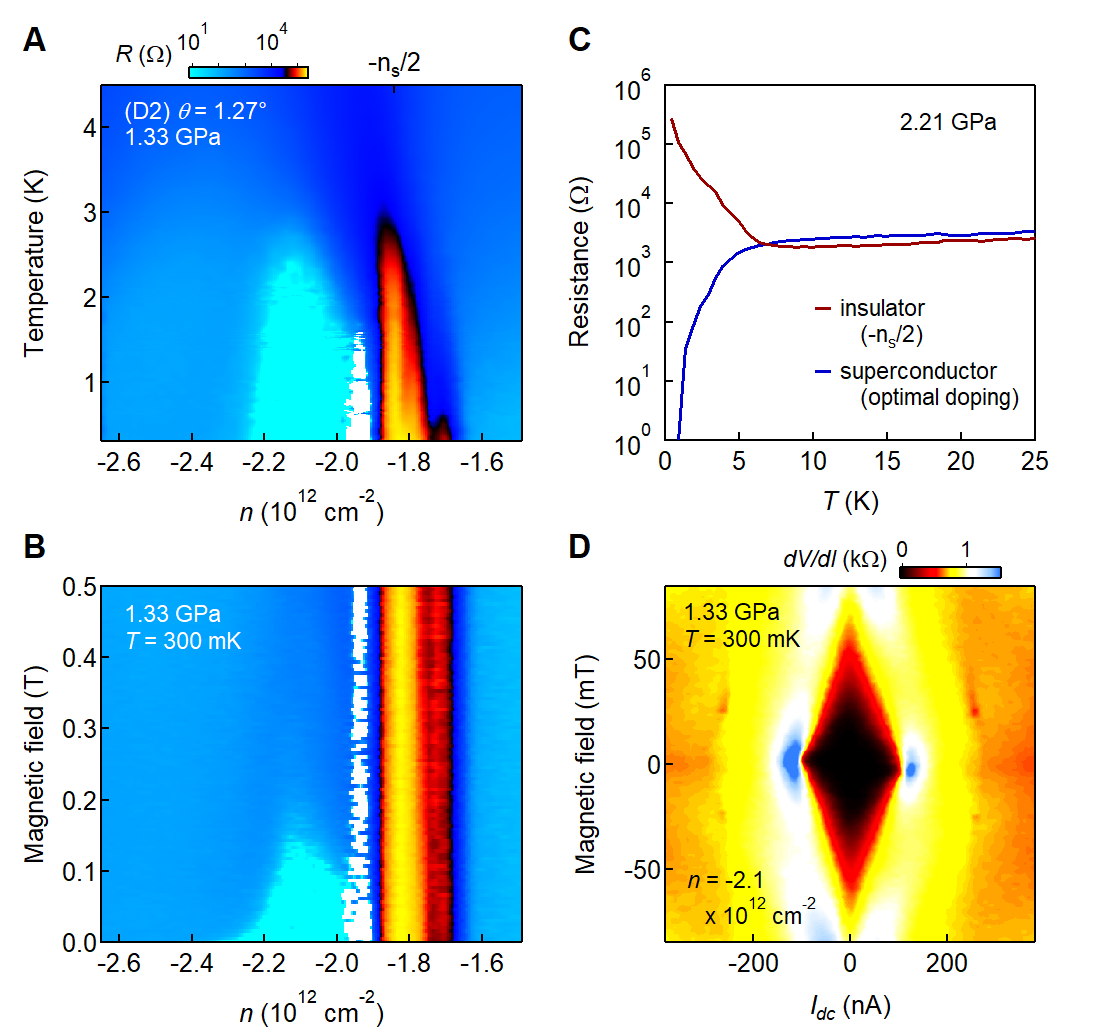} 
\caption{\textbf{Phase diagram of tBLG under pressure.}
(\textbf{A}) Resistance of device D2 (1.27$^{\circ}$) over a small range of carrier density near $-n_s/2$ versus $T$. An insulating phase at $-n_s/2$ neighbors a superconducting pocket at slightly larger hole doping. An apparent metallic phase separates the two~\cite{SI}, but is obscured by a region of artificially negative resistance in the contacts used for this measurement (colored in white).
(\textbf{B}) Similar map as a function of $B$.
(\textbf{C}) Resistance as a function of $T$ at 2.21 GPa at $-n_s/2$ (red) and at optimal doping of the superconductor (blue).
(\textbf{D}) Map of $dV/dI$ versus $I_{dc}$ and $B$ at $n = -2.1 \times 10^{12}$ cm$^{-2}$, $T$ = 300 mK, and 1.33 GPa. The map is acquired using different contacts from panels (A) and (B), and in particular exhibits a lower upper critical field $H_{c2}$. 
}
\label{fig:3}
\end{figure*}

\bigskip
\bigskip
\noindent\textbf{Bandwidth tuning with pressure}\\
The width of the flat bands in tBLG is determined by an interplay between the momentum-space mismatch of the Dirac cones between the graphene layers (set by the twist angle) and the strength of the layer hybridization (set by the interlayer spacing). The graphene interlayer spacing can be decreased by applying hydrostatic pressure~\cite{Yankowitz2018}, while leaving the interlayer rotation fixed~\cite{SI}. Pressure can therefore theoretically be used to achieve the flat band condition at arbitrary twist angle, relaxing the need for precise angle tuning~\cite{Carr2018,Chittari2018}. Furthermore, inducing the flat band at higher twist angle has been proposed as route towards increasing the energy scale of the superconductor\cite{Cao2018b,Carr2018,Chittari2018}.

Fig. ~\ref{fig:2}A shows the conductance $G$ versus density for a device, D2, with twist angle $\theta \approx$ 1.27$^{\circ}$. The gray curve in Fig.~\ref{fig:2}A shows the ambient-pressure response. Strongly insulating states appear at full filling of the moir\'e unit cell $\pm n_s$, indicating the presence of an isolated low-energy band. However, only very weakly insulating states (conductance minima) are observed at $\pm n_s/2$ and around $\pm 3n_s/4$, and no evidence of superconductivity is present, suggesting at this angle the low energy band does not support strong correlations. The blue curve shows the conductance of the same device under 2.21 GPa of hydrostatic pressure. Insulating states at several rational fillings of the moir\'e unit cell become evident --- most notably at $\pm n_s/2$ and $+3n_s/4$, as well as more weakly at $+n_s/4$ and $-3n_s/4$. Fig.~\ref{fig:2}B plots the conductance of the device at $-n_s/2$ as a function of temperature for the three values of pressure we applied, illustrating the crossover from metallic to insulating behavior under pressure and consistent with pressure-induced bandwidth tuning. 

In addition to strong insulating phases, we also observe a pressure-induced emergence of superconductivity. For hole doping slightly beyond $-n_s/2$, the device exhibits metallic temperature dependence under ambient pressure but superconducting behavior at high pressure, with the resistance rapidly dropping to the experimental noise floor of $\sim$10 $\Omega$ (Fig.~\ref{fig:2}C). In the pressure range that we study, the insulating gaps (measured by thermal activation, Fig.~\ref{fig:2}D) and $T_c$ of the superconductor (Fig.~\ref{fig:2}E) vary non-monotonically with pressure, with both reaching their highest measured values at 1.33 GPa. We also observe the onset of electron-type superconductivity for electron-doping just larger than $+n_s/2$, as evidenced by a sharp drop in the device magnetoresistance around $B$ = 0~\cite{SI}. However it appears to have a much lower $T_c$ than its hole-doping counterpart, preventing detailed study in pressure experiments where our base temperature was limited to 300 mK.

Recent bandstructure calculations indicate that the relation between bandwidth and pressure depends on the twist angle~\cite{Carr2018,Chittari2018}. For a 1.27$^{\circ}$ tBLG the minimum bandwidth is theoretically predicted to be in the range of 1.3 GPa to 1.5 GPa~\cite{Carr2018,Chittari2018}, in remarkably good agreement with the pressure value where we observed maximum $T_c$. We note that the largest measured $T_c$ we induce with pressure is $\sim$3 K, nearly an order of magnitude larger than observed in Device D1 and roughly a factor 2 larger than reported by Cao \textit{et al.}~\cite{Cao2018b}. This relative increase in $T_c$ could result from an increase in the Coulomb interaction energy scale resulting from the reduced moir\'e wavelength at the larger twist angle of this device~\cite{Cao2018b,Carr2018,Chittari2018}, but may also relate to differences in sample disorder.

\begin{figure*}[ht]
\includegraphics[width=6.9 in]{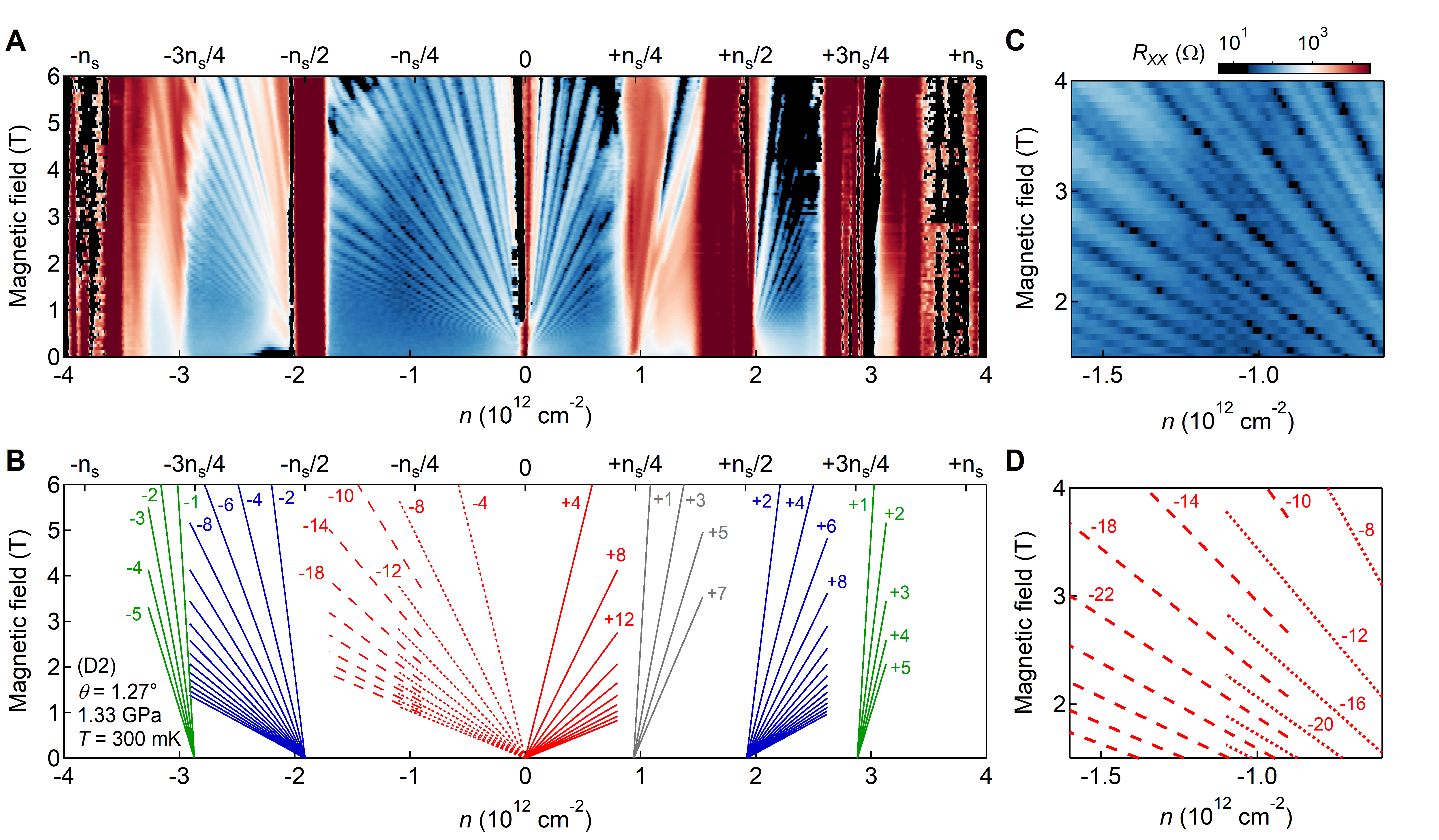} 
\caption{\textbf{Quantum oscillations in flat-band twisted BLG.}
(\textbf{A}) Landau fan diagram of device D2 (1.27$^{\circ}$) at 1.33 GPa up to full-filling of the moir\'e unit cell at $T$ = 300 mK. Quantum oscillations emerge from the CNP with dominant degeneracy sequence of $\nu = \pm 4, \pm 8, \pm 12,...$ at low field. Separate sequences of quantum oscillations emerge from $+n_s/4$ with dominant sequence of $\nu = +1, +3, +5,...$, from $\pm n_s/2$ with dominant sequence of $\nu = \pm 2, \pm 4, \pm 6,...$, and from $\pm 3n_s/4$ with dominant sequence of $\nu = \pm 1, \pm 2, \pm 3,...$. Regions of negative measured voltage are set to zero resistance for clarity, most prominently affecting the high-field region of the map between $+n_s/2$ and $+3n_s/4$. 
(\textbf{B}) Schematic Landau level structure corresponding to the observations in (A). Only the Landau levels persisting to the lowest fields are plotted, while by 6 T states at all filling factors are observed.
(\textbf{C}) Zoom-in of (A) around $-n_s/4$. 
(\textbf{D}) Schematic Landau level structure corresponding to the observations in (C). The dominant degeneracy sequence evolves smoothly from $\nu = -4, -8, -12,...$ at low density to $\nu = -10, -14, -18,...$ at high density, switching around $-n_s/4$.
}
\label{fig:4}
\end{figure*}

Figs.~\ref{fig:3}A and B plot the device resistance around $-n_s/2$ as a function of temperature $T$ and magnetic field $B$, respectively, under a pressure of 1.33 GPa. We observe a state with strongly insulating temperature dependence at $-n_s/2$, with a pocket of superconductivity at slightly larger hole doping and metallic behavior at slightly smaller hole doping. Our results differ from those of our device D1 and of the devices reported in Ref.~\onlinecite{Cao2018b} in several ways. First, the device resistance in Fig.~\ref{fig:3}A grows quickly as the temperature is lowered, while in prior devices it drops towards zero (e.g., for negative values of $D$ for device D1). Second, we observe a pocket of superconductivity \textit{only} for  $n < -n_s/2$, whereas prior devices additionally exhibit superconductivity for $n > -n_s/2$. Third, we find evidence for a metallic phase separating the superconducting and insulating phases. While this is partially obscured in Fig.~\ref{fig:3}A by an anomalous region of apparent negative resistance in this region arising from a measurement artifact (colored in white), measurements using other contacts (as well similar measurements at 2.21 GPa) exhibit more clear evidence of the metallic phase~\cite{SI}.

Fig.~\ref{fig:3}C shows a comparison of the resistance versus temperature of the insulating (red) and superconducting (blue) state, measured at 2.21 GPa. The two phases onset at remarkably similar temperatures with the resistance in both cases diverging at around 5~K from the high temperature metallic behavior. Similar behavior is also observed in device D1, where additionally we found that similar critical currents quench each phase to the normal state resistance~\cite{SI}. These observations suggest that the insulating and superconducting phases share similar energy scales, constraining models in which the superconductivity arises as a daughter-state of the insulator.

The lack of strong magnetoresistance oscillations in Fig.~\ref{fig:3}B suggests that this sample is highly homogeneous. To confirm this, we plot $dV/dI$ as a function of $B$ and $I_{dc}$ in Fig.~\ref{fig:3}D, and find that $I_c$ decreases roughly linearly with $B$ and, unlike device D1, does not exhibit quantum interference patterns associated with junction-limited superconductivity. We have additionally measured the phase diagram of this device as a function of $D$, and do not observe a significant displacement field dependence (in particular there is a robust insulating state at $-n_s/2$ for all $D$)~\cite{SI}. We interpret these observations to indicate that this device is less disordered than those previously reported, suggesting that details of the associated superconducting and insulating response may more faithfully represent the disorder-free phase diagram. The reasons for the reduced disorder in this sample are not fully understood. This may be emblematic of larger twist angle devices where the combination of smaller moir\'e period and applied pressure minimize the contribution of spatial inhomogeneity. However, owing to the limited sample size we also can not rule out random sample-to-sample variation. A systematic study of the interplay between twist angle and pressure --- preferably in a single device~\cite{Yankowitz2018,Ribeiro2018} --- will be needed to resolve these issues. 

\bigskip
\noindent\textbf{Quantum oscillations and new Fermi surfaces}\\
The high degree of structural and charge homogeneity in our samples further allows high resolution measurements of magnetoresistance oscillations associated with cyclotron motion of electrons. Quantum oscillations at low magnetic fields give detailed information about electronic band structure, as their periodicity can be used to infer the areal size of the Fermi surface. Moreover, their degeneracy reflects the presence of spin, valley, and layer degrees of freedom. Fig.~\ref{fig:4}A shows magnetoresistance data from device D2 at 1.33 GPa. Several sets of seemingly independent Landau fans are observed, indicated schematically in Fig.~\ref{fig:4}B. In contrast to devices with larger twist angle, none of the quantum oscillations show $D$-tuned Landau level crossings~\cite{SI}. Near the CNP, we observe a 4-fold degenerate sequence of quantum oscillations, with dominant minima at $\nu = \pm 4, \pm 8, \pm 12,...$ at low magnetic field, where $\nu=n h/(eB)$ is the Landau level filling factor relative to charge neutrality. 

At $+n_s/4$, $\pm n_s/2$, and $\pm 3n_s/4$, the carrier density extracted from the Hall effect approaches zero~\cite{SI}, indicating the formation of new, small Fermi surfaces. These fillings also spawn independent series of quantum oscillations. At $\pm n_s/2$, oscillations clearly exhibit 2-fold degeneracy at very low fields, suggesting that the combined spin- and valley- degeneracy is partially lifted~\cite{Cao2018b}. In contrast, the sequences of quantum oscillations emerging from $\pm 3n_s/4$ exhibit no additional degeneracy, suggesting that all degeneracies are lifted in these Fermi surfaces. 

The sequence emerging from $+n_s/4$ appears to be 2-fold degenerate but, interestingly, is odd-dominant with primary oscillations observed at $\nu = +1, +3, +5,...$, where $\nu$ is defined relative to $+n_s/4$. While neither a resistive state nor quantum oscillations are observed originating from $-n_s/4$, we observe a shift in the dominant filling sequence for the CNP fan when the carrier density reaches $-n_s/4$. Around this density, the best pronounced oscillations transition from a $\nu = -4, -8, -12,...$ sequence near the CNP to $\nu = -10, -14, -18,...$ for $n<-n_s/4$. Fig.~\ref{fig:4}C-D show a detailed view of this transition highlighting the crossover of the dominant sequence. 

Some of our observations are likely accounted for by the single particle band structure of tBLG. However, currently available models predict eightfold degeneracy near the CNP, arising from spin, valley, and layer degrees of freedom~\cite{Yuan2018}. We note that our observed CNP sequence is identical to that in Bernal stacked bilayer graphene, where interlayer tunneling leads to hybridization. Quadratic band touching reminiscent of that system does feature in several recent models of tBLG flat bands~\cite{Song2018,Hejazi2018}; however, these models also feature additional Dirac point band crossings at low energy.  Massless Dirac points may play a role in phase shifts observed in the quantum oscillations near $\pm n_s/4$: the $1,3,5...$ and $10,14,18...$ sequences are consistent with quantum oscillations of a single massless Dirac cone with $\pi$-Berry phase, with two-fold and four-fold degeneracy, respectively. Alternatively, these phase shifts can arise when cyclotron mass $m^*$ becomes comparable to the bare electron mass $m_e$, leading to a coincidence between Zeeman split Landau levels in different orbitals. Cyclotron mass measurements may be able to distinguish these two band-structure driven scenarios.

In contrast to the features near the CNP, a single-particle model fails to predict the emergence of new sequences of quantum oscillations emerging from gapped states at commensurate fillings of the moir\'e unit cell. Furthermore, for each of the commensurate fillings, the associated quantum oscillations disperse only in a single direction for each carrier type (away from the CNP) and abruptly terminate at the next commensurate filling. In contrast, a simple band structure model with gapped bands at commensurate filling would lead to quantum oscillations emerging in both directions, raising questions regarding the detailed structure of the interaction-modified bands.

\bigskip
\noindent\textbf{Isospin ordering of the correlated insulators}\\
Our observation that superconductivity appears only at densities coincident with lower-degeneracy quantum oscillations near half-filling --- but not near insulating states at quarter- and three-quarter filling --- suggests that the nature of symmetry breaking may be integral to superconductivity. A wide array of candidate states have already been proposed in the literature to describe the correlated insulating states, including antiferromagnetic Mott insulators, charge density waves, Wigner crystals, and isospin ferromagnetic band  insulators~\cite{Xu2018, Po2018, Yuan2018, Roy2018, Baskaran2018, Isobe2018, Padhi2018, Ochi2018}. Parallel magnetic field measurements offer a simple probe of isospin physics, as a primary effect of $B_\parallel$ is to increase the Zeeman energy. Under applied $B_\parallel$, the gap of a spin-unpolarized (valley-polarized) ground state should decrease linearly, while the gap is expected to increase linearly, or remain unchanged, for spin-polarized (valley-unpolarized) ground state ordering.

\begin{figure}[t]
\includegraphics[width=2.3in]{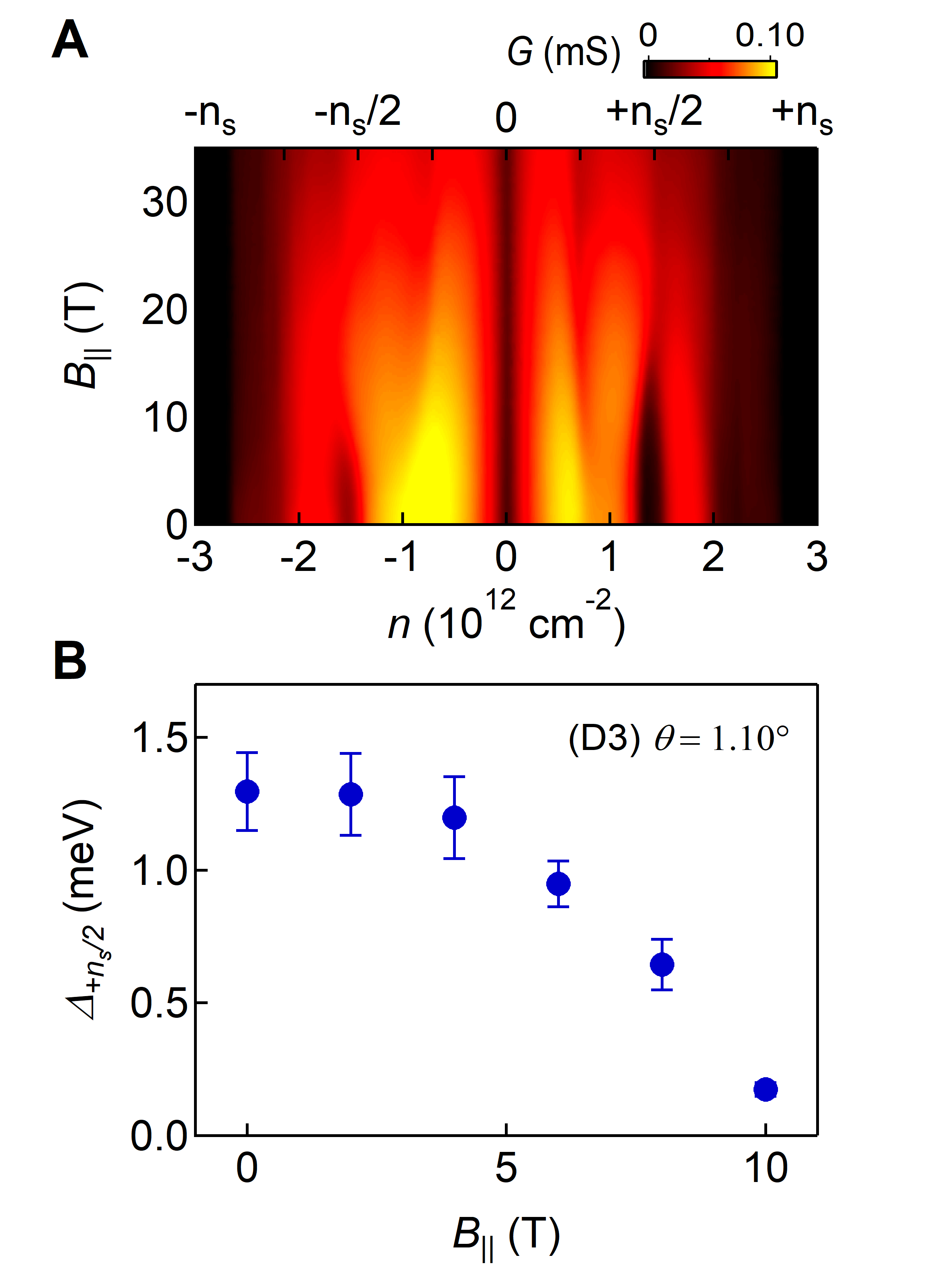} 
\caption{\textbf{Parallel field dependence of correlated insulating states.}
(\textbf{A}) Conductance of device D3 (1.10$^{\circ}$) as a function of carrier density and $B_\parallel$ at $T$ = 300 mK. Correlated insulating states at $\pm n_s/2$ are less insulating at large $B_\parallel$, while states at $\pm n_s/4$ are more insulating. The conductance of states at $\pm 3n_s/4$ and at the CNP are roughly independent of $B_\parallel$.
(\textbf{B}) Energy gap of the correlated insulating state at $+n_s/2$ at various $B_\parallel$ measured by thermal activation. Error bars in the gaps represent the uncertainty arising from determining the linear (thermally activated) regime for the fit.
}
\label{fig:5}
\end{figure}

Fig.~\ref{fig:5}A shows the parallel magnetic-field response of the conductance of device D3, with twist angle $\theta \approx$ 1.10$^{\circ}$. The conductance at zero magnetic field exhibits minima at the CNP, $\pm n_s/2$, and $+3n_s/4$ band filling, along with more weakly developed features at $-3n_s/4$ and $+n_s/4$. At $\pm n_s/2$ filling, the conductance minimum fades with increasing $B_\parallel$, while the opposite behavior is observed at $\pm n_s/4$. The $\pm 3n_s/4$ states are insensitive to magnetic field. Qualitatively, our observations suggest that the $\pm n_{s}/2$ insulators are spin-unpolarized, whereas both the $\pm n_{s}/4$ and $\pm 3n_{s}/4$ are spin-polarized. 

We explore this picture quantitatively using the thermal activation gap of the $+n_s/2$ state --- the only commensurate insulator that shows clear thermal activation in this device --- as a function of $B_\parallel$ (Fig.~\ref{fig:5}B). Unexpectedly, it closes non-linearly with $B_\parallel$, inconsistent with either full spin- or valley-polarization. The approximately parabolic decrease is also not consistent with antiferromagnetically aligned spins in opposite valleys, which would lead to an increasing gap at larger $B_\parallel$. Our finding suggests that isospin ordering alone, without additional symmetry breaking that lifts the degeneracy of the Dirac band touching, may be insufficient to understand the nature of the half-filling insulators.

\bigskip
\noindent\textbf{Discussion}\\
Accurate modeling of twisted bilayer graphene is complicated by the the large size of the moir\'e unit cell, which prevents reliable band structure calculations.  Moreover, emerging literature suggests a wide variety of plausible mechanisms and ground states for both the superconducting and insulating states. While many of these models predict unconventional all-electronic pairing mechanisms~\cite{Guo2018,Xu2018,Roy2018,Baskaran2018,Huang2018,Zhang2018b,Ray2018,Liu2018,Fidrysiak2018,Rademaker2018,Isobe2018,Kennes2018,You2018,Gonzalez2018,Su2018,Sherkunov2018}, others are either explicitly or implicitly consistent with conventional phonon-mediated superconductivity~\cite{Dodaro2018,Peltonen2018,Wu2018b,Ochi2018,Lian2018,Laksono2018}. Our data, taken together, highlight a number of constraints any successful theory of tBLG should satisfy, which we summarize here.  

First, our experiments confirm the basic notion that correlation physics in tBLG arises from the interplay of angular misalignment and interlayer tunneling~\cite{Bistritzer2011}.  We observe insulating and superconducting states in the same regime of angles as Cao \textit{et al.}~\cite{Cao2018a,Cao2018b} at ambient pressure, and our finite-pressure experiments show the anticipated trend, with flat-band physics appearing at larger angles for larger applied pressure. The small and likely extrinsic effect of displacement field is similarly consistent with expectations of strong interlayer hybridization. 

However, our data reveal several important new details of the resulting correlated states themselves. Most importantly, our data suggest that the intrinsic domain of superconductivity in this system is restricted to a narrow range of charge density at higher absolute density --- but not lower --- than half-filling (i.e. $|n| > |\pm n_s/2|$). Superconductivity occurs for both electron- and hole-doping, and is nearly adjacent to a strongly insulating state. The domain of superconductivity coincides with that of two-fold degenerate quantum oscillations, indicating that a small Fermi surface with reduced degeneracy is nucleated at $n_s/2$. Although we cannot exclude superconducting states with transitions at temperatures lower than our experimental base temperature, we fail to observe superconductivity near other commensurate fillings, including several which host similar insulating states. This suggests that the nature of the Fermi surface nucleated $\pm n_s/2$ may be essential to the onset of nearby superconductivity, although whether this is via a purely electronic mechanism, an enhancement of density of states that favor a phonon-mediated mechanism, or some mechanism yet to be proposed is unknown. 

Finally, our results suggest that future work should focus on improving the spatial homogeneity of the tBLG moir\'e pattern. In this study, we realized a small electronic bandwidth in a device with a smaller moir\'e period by applying pressure. Local strains in the graphene lattice result in smaller fluctuations of the moir\'e period as the overall moir\'e period is reduced~\cite{Yoo2018}, therefore we anticipate that applying higher pressure to a device with even larger twist angle could result in further improvements to the device homogeneity. Additionally, the smaller moir\'e period is anticipated to drive stronger Coulomb repulsion, potentially leading to larger energy scales for the superconducting and correlated insulating phases~\cite{Cao2018b,Carr2018,Chittari2018}. Reliable fabrication of highly homogeneous samples will be critical for further progress in understanding the mechanisms driving the correlated phases observed in these systems, especially in experiments performed without the use of pressure.

\subsubsection*{Acknowledgements}

The authors acknowledge experimental assistance from Jiacheng Zhu and Haoxin Zhou, and helpful discussions with Dan Shahar, Andrew Millis, Oskar Vafek, Mike Zaletel, Leon Balents, Cenke Xu, Andrei Bernevig, Liang Fu, Mikito Koshino, and Pilkyung Moon.

\subsubsection*{Funding}
Work at both Columbia and UCSB was funded by the Army Research Office under W911NF-17-1-0323. AFY and CRD separately acknowledge the support of the David and Lucile Packard Foundation. A portion of this work was performed at the National High Magnetic Field Laboratory, which is supported by the National Science Foundation Cooperative Agreement No. DMR-1644779 and the State of Florida. K.W. and T.T. acknowledge support from the Elemental Strategy Initiative conducted by the MEXT, Japan, and JSPS KAKENHI grant no. JP15K21722. 

\subsubsection*{Author contributions}
M.Y. and S.C. fabricated the devices. M.Y., S.C. and H.P. performed the measurements and analyzed the data. D.G. loaded the pressure cell. K.W. and T.T. grew the hBN crystals. A.F.Y. and C.R.D. advised on the experiments. The manuscript was written with input from all authors. 

\subsubsection*{Competing interests}
The authors declare no competing financial interests.

\bibliography{references}

\begin{thebibliography}{59}%
\makeatletter
\providecommand \@ifxundefined [1]{%
 \@ifx{#1\undefined}
}%
\providecommand \@ifnum [1]{%
 \ifnum #1\expandafter \@firstoftwo
 \else \expandafter \@secondoftwo
 \fi
}%
\providecommand \@ifx [1]{%
 \ifx #1\expandafter \@firstoftwo
 \else \expandafter \@secondoftwo
 \fi
}%
\providecommand \natexlab [1]{#1}%
\providecommand \enquote  [1]{``#1''}%
\providecommand \bibnamefont  [1]{#1}%
\providecommand \bibfnamefont [1]{#1}%
\providecommand \citenamefont [1]{#1}%
\providecommand \href@noop [0]{\@secondoftwo}%
\providecommand \href [0]{\begingroup \@sanitize@url \@href}%
\providecommand \@href[1]{\@@startlink{#1}\@@href}%
\providecommand \@@href[1]{\endgroup#1\@@endlink}%
\providecommand \@sanitize@url [0]{\catcode `\\12\catcode `\$12\catcode
  `\&12\catcode `\#12\catcode `\^12\catcode `\_12\catcode `\%12\relax}%
\providecommand \@@startlink[1]{}%
\providecommand \@@endlink[0]{}%
\providecommand \url  [0]{\begingroup\@sanitize@url \@url }%
\providecommand \@url [1]{\endgroup\@href {#1}{\urlprefix }}%
\providecommand \urlprefix  [0]{URL }%
\providecommand \Eprint [0]{\href }%
\providecommand \doibase [0]{http://dx.doi.org/}%
\providecommand \selectlanguage [0]{\@gobble}%
\providecommand \bibinfo  [0]{\@secondoftwo}%
\providecommand \bibfield  [0]{\@secondoftwo}%
\providecommand \translation [1]{[#1]}%
\providecommand \BibitemOpen [0]{}%
\providecommand \bibitemStop [0]{}%
\providecommand \bibitemNoStop [0]{.\EOS\space}%
\providecommand \EOS [0]{\spacefactor3000\relax}%
\providecommand \BibitemShut  [1]{\csname bibitem#1\endcsname}%
\let\auto@bib@innerbib\@empty
\bibitem [{\citenamefont {Bistritzer}\ and\ \citenamefont
  {MacDonald}(2011)}]{Bistritzer2011}%
  \BibitemOpen
  \bibfield  {author} {\bibinfo {author} {\bibfnamefont {R.}~\bibnamefont
  {Bistritzer}}\ and\ \bibinfo {author} {\bibfnamefont {A.~H.}\ \bibnamefont
  {MacDonald}},\ }\href@noop {} {\bibfield  {journal} {\bibinfo  {journal}
  {Proceedings of the National Academy of Sciences}\ }\textbf {\bibinfo
  {volume} {108}},\ \bibinfo {pages} {12233} (\bibinfo {year}
  {2011})}\BibitemShut {NoStop}%
\bibitem [{\citenamefont {Chen}\ \emph {et~al.}(2018)\citenamefont {Chen},
  \citenamefont {Jiang}, \citenamefont {Wu}, \citenamefont {Lv}, \citenamefont
  {Li}, \citenamefont {Watanabe}, \citenamefont {Taniguchi}, \citenamefont
  {Shi}, \citenamefont {Zhang},\ and\ \citenamefont {Wang}}]{Chen2018}%
  \BibitemOpen
  \bibfield  {author} {\bibinfo {author} {\bibfnamefont {G.}~\bibnamefont
  {Chen}}, \bibinfo {author} {\bibfnamefont {L.}~\bibnamefont {Jiang}},
  \bibinfo {author} {\bibfnamefont {S.}~\bibnamefont {Wu}}, \bibinfo {author}
  {\bibfnamefont {B.}~\bibnamefont {Lv}}, \bibinfo {author} {\bibfnamefont
  {H.}~\bibnamefont {Li}}, \bibinfo {author} {\bibfnamefont {K.}~\bibnamefont
  {Watanabe}}, \bibinfo {author} {\bibfnamefont {T.}~\bibnamefont {Taniguchi}},
  \bibinfo {author} {\bibfnamefont {Z.}~\bibnamefont {Shi}}, \bibinfo {author}
  {\bibfnamefont {Y.}~\bibnamefont {Zhang}}, \ and\ \bibinfo {author}
  {\bibfnamefont {F.}~\bibnamefont {Wang}},\ }\href@noop {} {\bibfield
  {journal} {\bibinfo  {journal} {arXiv:1803.01985}\ } (\bibinfo {year}
  {2018})}\BibitemShut {NoStop}%
\bibitem [{\citenamefont {Wu}\ \emph {et~al.}(2018{\natexlab{a}})\citenamefont
  {Wu}, \citenamefont {Lovorn}, \citenamefont {Tutuc},\ and\ \citenamefont
  {MacDonald}}]{Wu2018a}%
  \BibitemOpen
  \bibfield  {author} {\bibinfo {author} {\bibfnamefont {F.}~\bibnamefont
  {Wu}}, \bibinfo {author} {\bibfnamefont {T.}~\bibnamefont {Lovorn}}, \bibinfo
  {author} {\bibfnamefont {E.}~\bibnamefont {Tutuc}}, \ and\ \bibinfo {author}
  {\bibfnamefont {A.~H.}\ \bibnamefont {MacDonald}},\ }\href@noop {} {\bibfield
   {journal} {\bibinfo  {journal} {Physical Review Letters}\ }\textbf {\bibinfo
  {volume} {121}},\ \bibinfo {pages} {026402} (\bibinfo {year}
  {2018}{\natexlab{a}})}\BibitemShut {NoStop}%
\bibitem [{\citenamefont {Zhang}\ \emph {et~al.}(2018)\citenamefont {Zhang},
  \citenamefont {Mao}, \citenamefont {Cao}, \citenamefont {Jarillo-Herrero},\
  and\ \citenamefont {Senthil}}]{Zhang2018}%
  \BibitemOpen
  \bibfield  {author} {\bibinfo {author} {\bibfnamefont {Y.-H.}\ \bibnamefont
  {Zhang}}, \bibinfo {author} {\bibfnamefont {D.}~\bibnamefont {Mao}}, \bibinfo
  {author} {\bibfnamefont {Y.}~\bibnamefont {Cao}}, \bibinfo {author}
  {\bibfnamefont {P.}~\bibnamefont {Jarillo-Herrero}}, \ and\ \bibinfo {author}
  {\bibfnamefont {T.}~\bibnamefont {Senthil}},\ }\href@noop {} {\bibfield
  {journal} {\bibinfo  {journal} {arXiv:1805.08232}\ } (\bibinfo {year}
  {2018})}\BibitemShut {NoStop}%
\bibitem [{\citenamefont {Cao}\ \emph {et~al.}(2018{\natexlab{a}})\citenamefont
  {Cao}, \citenamefont {Fatemi}, \citenamefont {Demir}, \citenamefont {Fang},
  \citenamefont {Tomarken}, \citenamefont {Luo}, \citenamefont
  {Sanchez-Yamagishi}, \citenamefont {Watanabe}, \citenamefont {Taniguchi},
  \citenamefont {Kaxiras}, \citenamefont {Ashoori},\ and\ \citenamefont
  {Jarillo-Herrero}}]{Cao2018a}%
  \BibitemOpen
  \bibfield  {author} {\bibinfo {author} {\bibfnamefont {Y.}~\bibnamefont
  {Cao}}, \bibinfo {author} {\bibfnamefont {V.}~\bibnamefont {Fatemi}},
  \bibinfo {author} {\bibfnamefont {A.}~\bibnamefont {Demir}}, \bibinfo
  {author} {\bibfnamefont {S.}~\bibnamefont {Fang}}, \bibinfo {author}
  {\bibfnamefont {S.~L.}\ \bibnamefont {Tomarken}}, \bibinfo {author}
  {\bibfnamefont {J.~Y.}\ \bibnamefont {Luo}}, \bibinfo {author} {\bibfnamefont
  {J.~D.}\ \bibnamefont {Sanchez-Yamagishi}}, \bibinfo {author} {\bibfnamefont
  {K.}~\bibnamefont {Watanabe}}, \bibinfo {author} {\bibfnamefont
  {T.}~\bibnamefont {Taniguchi}}, \bibinfo {author} {\bibfnamefont
  {E.}~\bibnamefont {Kaxiras}}, \bibinfo {author} {\bibfnamefont {R.~C.}\
  \bibnamefont {Ashoori}}, \ and\ \bibinfo {author} {\bibfnamefont
  {P.}~\bibnamefont {Jarillo-Herrero}},\ }\href@noop {} {\bibfield  {journal}
  {\bibinfo  {journal} {Nature}\ }\textbf {\bibinfo {volume} {556}},\ \bibinfo
  {pages} {80} (\bibinfo {year} {2018}{\natexlab{a}})}\BibitemShut {NoStop}%
\bibitem [{\citenamefont {Cao}\ \emph {et~al.}(2018{\natexlab{b}})\citenamefont
  {Cao}, \citenamefont {Fatemi}, \citenamefont {Fang}, \citenamefont
  {Watanabe}, \citenamefont {Taniguchi}, \citenamefont {Kaxiras},\ and\
  \citenamefont {Jarillo-Herrero}}]{Cao2018b}%
  \BibitemOpen
  \bibfield  {author} {\bibinfo {author} {\bibfnamefont {Y.}~\bibnamefont
  {Cao}}, \bibinfo {author} {\bibfnamefont {V.}~\bibnamefont {Fatemi}},
  \bibinfo {author} {\bibfnamefont {S.}~\bibnamefont {Fang}}, \bibinfo {author}
  {\bibfnamefont {K.}~\bibnamefont {Watanabe}}, \bibinfo {author}
  {\bibfnamefont {T.}~\bibnamefont {Taniguchi}}, \bibinfo {author}
  {\bibfnamefont {E.}~\bibnamefont {Kaxiras}}, \ and\ \bibinfo {author}
  {\bibfnamefont {P.}~\bibnamefont {Jarillo-Herrero}},\ }\href@noop {}
  {\bibfield  {journal} {\bibinfo  {journal} {Nature}\ }\textbf {\bibinfo
  {volume} {556}},\ \bibinfo {pages} {43} (\bibinfo {year}
  {2018}{\natexlab{b}})}\BibitemShut {NoStop}%
\bibitem [{\citenamefont {Carr}\ \emph {et~al.}(2018)\citenamefont {Carr},
  \citenamefont {Fang}, \citenamefont {Jarillo-Herrero},\ and\ \citenamefont
  {Kaxiras}}]{Carr2018}%
  \BibitemOpen
  \bibfield  {author} {\bibinfo {author} {\bibfnamefont {S.}~\bibnamefont
  {Carr}}, \bibinfo {author} {\bibfnamefont {S.}~\bibnamefont {Fang}}, \bibinfo
  {author} {\bibfnamefont {P.}~\bibnamefont {Jarillo-Herrero}}, \ and\ \bibinfo
  {author} {\bibfnamefont {E.}~\bibnamefont {Kaxiras}},\ }\href@noop {}
  {\bibfield  {journal} {\bibinfo  {journal} {arXiv:1806.05078}\ } (\bibinfo
  {year} {2018})}\BibitemShut {NoStop}%
\bibitem [{\citenamefont {Chittari}\ \emph {et~al.}(2018)\citenamefont
  {Chittari}, \citenamefont {Leconte}, \citenamefont {Javvaji},\ and\
  \citenamefont {Jung}}]{Chittari2018}%
  \BibitemOpen
  \bibfield  {author} {\bibinfo {author} {\bibfnamefont {B.~L.}\ \bibnamefont
  {Chittari}}, \bibinfo {author} {\bibfnamefont {N.}~\bibnamefont {Leconte}},
  \bibinfo {author} {\bibfnamefont {S.}~\bibnamefont {Javvaji}}, \ and\
  \bibinfo {author} {\bibfnamefont {J.}~\bibnamefont {Jung}},\ }\href@noop {}
  {\bibfield  {journal} {\bibinfo  {journal} {arXiv:1808.00104}\ } (\bibinfo
  {year} {2018})}\BibitemShut {NoStop}%
\bibitem [{\citenamefont {Trambly De~Laissardi\`ere}\ \emph
  {et~al.}(2016)\citenamefont {Trambly De~Laissardi\`ere}, \citenamefont
  {Namarvar}, \citenamefont {Mayou},\ and\ \citenamefont
  {Magaud}}]{Laissardiere2016}%
  \BibitemOpen
  \bibfield  {author} {\bibinfo {author} {\bibfnamefont {G.}~\bibnamefont
  {Trambly De~Laissardi\`ere}}, \bibinfo {author} {\bibfnamefont {O.~F.}\
  \bibnamefont {Namarvar}}, \bibinfo {author} {\bibfnamefont {D.}~\bibnamefont
  {Mayou}}, \ and\ \bibinfo {author} {\bibfnamefont {L.}~\bibnamefont
  {Magaud}},\ }\href@noop {} {\bibfield  {journal} {\bibinfo  {journal}
  {Physical Review B}\ }\textbf {\bibinfo {volume} {93}},\ \bibinfo {pages}
  {235135} (\bibinfo {year} {2016})}\BibitemShut {NoStop}%
\bibitem [{\citenamefont {Zibrov}\ \emph {et~al.}(2017)\citenamefont {Zibrov},
  \citenamefont {Kometter}, \citenamefont {Zhou}, \citenamefont {Spanton},
  \citenamefont {Taniguchi}, \citenamefont {Watanabe}, \citenamefont
  {Zaletel},\ and\ \citenamefont {Young}}]{Zibrov2017}%
  \BibitemOpen
  \bibfield  {author} {\bibinfo {author} {\bibfnamefont {A.~A.}\ \bibnamefont
  {Zibrov}}, \bibinfo {author} {\bibfnamefont {C.}~\bibnamefont {Kometter}},
  \bibinfo {author} {\bibfnamefont {H.}~\bibnamefont {Zhou}}, \bibinfo {author}
  {\bibfnamefont {E.~M.}\ \bibnamefont {Spanton}}, \bibinfo {author}
  {\bibfnamefont {T.}~\bibnamefont {Taniguchi}}, \bibinfo {author}
  {\bibfnamefont {K.}~\bibnamefont {Watanabe}}, \bibinfo {author}
  {\bibfnamefont {M.~P.}\ \bibnamefont {Zaletel}}, \ and\ \bibinfo {author}
  {\bibfnamefont {A.~F.}\ \bibnamefont {Young}},\ }\href@noop {} {\bibfield
  {journal} {\bibinfo  {journal} {Nature}\ }\textbf {\bibinfo {volume} {549}},\
  \bibinfo {pages} {360} (\bibinfo {year} {2017})}\BibitemShut {NoStop}%
\bibitem [{\citenamefont {Yankowitz}\ \emph {et~al.}(2018)\citenamefont
  {Yankowitz}, \citenamefont {Jung}, \citenamefont {Laksono}, \citenamefont
  {Leconte}, \citenamefont {Chittari}, \citenamefont {Watanabe}, \citenamefont
  {Taniguchi}, \citenamefont {Adam}, \citenamefont {Graf},\ and\ \citenamefont
  {Dean}}]{Yankowitz2018}%
  \BibitemOpen
  \bibfield  {author} {\bibinfo {author} {\bibfnamefont {M.}~\bibnamefont
  {Yankowitz}}, \bibinfo {author} {\bibfnamefont {J.}~\bibnamefont {Jung}},
  \bibinfo {author} {\bibfnamefont {E.}~\bibnamefont {Laksono}}, \bibinfo
  {author} {\bibfnamefont {N.}~\bibnamefont {Leconte}}, \bibinfo {author}
  {\bibfnamefont {B.~L.}\ \bibnamefont {Chittari}}, \bibinfo {author}
  {\bibfnamefont {K.}~\bibnamefont {Watanabe}}, \bibinfo {author}
  {\bibfnamefont {T.}~\bibnamefont {Taniguchi}}, \bibinfo {author}
  {\bibfnamefont {S.}~\bibnamefont {Adam}}, \bibinfo {author} {\bibfnamefont
  {D.}~\bibnamefont {Graf}}, \ and\ \bibinfo {author} {\bibfnamefont {C.~R.}\
  \bibnamefont {Dean}},\ }\href@noop {} {\bibfield  {journal} {\bibinfo
  {journal} {Nature}\ }\textbf {\bibinfo {volume} {557}},\ \bibinfo {pages}
  {404} (\bibinfo {year} {2018})}\BibitemShut {NoStop}%
\bibitem [{\citenamefont {Kim}\ \emph {et~al.}(2016)\citenamefont {Kim},
  \citenamefont {Yankowitz}, \citenamefont {Fallahazad}, \citenamefont {Kang},
  \citenamefont {Movva}, \citenamefont {Huang}, \citenamefont {Larentis},
  \citenamefont {Corbet}, \citenamefont {Taniguchi}, \citenamefont {Watanabe},
  \citenamefont {Banerjee}, \citenamefont {LeRoy},\ and\ \citenamefont
  {Tutuc}}]{Kim2016}%
  \BibitemOpen
  \bibfield  {author} {\bibinfo {author} {\bibfnamefont {K.}~\bibnamefont
  {Kim}}, \bibinfo {author} {\bibfnamefont {M.}~\bibnamefont {Yankowitz}},
  \bibinfo {author} {\bibfnamefont {B.}~\bibnamefont {Fallahazad}}, \bibinfo
  {author} {\bibfnamefont {S.}~\bibnamefont {Kang}}, \bibinfo {author}
  {\bibfnamefont {H.~C.~P.}\ \bibnamefont {Movva}}, \bibinfo {author}
  {\bibfnamefont {S.}~\bibnamefont {Huang}}, \bibinfo {author} {\bibfnamefont
  {S.}~\bibnamefont {Larentis}}, \bibinfo {author} {\bibfnamefont {C.~M.}\
  \bibnamefont {Corbet}}, \bibinfo {author} {\bibfnamefont {T.}~\bibnamefont
  {Taniguchi}}, \bibinfo {author} {\bibfnamefont {K.}~\bibnamefont {Watanabe}},
  \bibinfo {author} {\bibfnamefont {S.~K.}\ \bibnamefont {Banerjee}}, \bibinfo
  {author} {\bibfnamefont {B.~J.}\ \bibnamefont {LeRoy}}, \ and\ \bibinfo
  {author} {\bibfnamefont {E.}~\bibnamefont {Tutuc}},\ }\href@noop {}
  {\bibfield  {journal} {\bibinfo  {journal} {Nano Letters}\ }\textbf {\bibinfo
  {volume} {16}},\ \bibinfo {pages} {1989} (\bibinfo {year}
  {2016})}\BibitemShut {NoStop}%
\bibitem [{\citenamefont {Kim}\ \emph {et~al.}(2017)\citenamefont {Kim},
  \citenamefont {DaSilva}, \citenamefont {Huang}, \citenamefont {Fallahazad},
  \citenamefont {Larentis}, \citenamefont {Taniguchi}, \citenamefont
  {Watanabe}, \citenamefont {LeRoy}, \citenamefont {MacDonald},\ and\
  \citenamefont {Tutuc}}]{Kim2017}%
  \BibitemOpen
  \bibfield  {author} {\bibinfo {author} {\bibfnamefont {K.}~\bibnamefont
  {Kim}}, \bibinfo {author} {\bibfnamefont {A.}~\bibnamefont {DaSilva}},
  \bibinfo {author} {\bibfnamefont {S.}~\bibnamefont {Huang}}, \bibinfo
  {author} {\bibfnamefont {B.}~\bibnamefont {Fallahazad}}, \bibinfo {author}
  {\bibfnamefont {S.}~\bibnamefont {Larentis}}, \bibinfo {author}
  {\bibfnamefont {T.}~\bibnamefont {Taniguchi}}, \bibinfo {author}
  {\bibfnamefont {K.}~\bibnamefont {Watanabe}}, \bibinfo {author}
  {\bibfnamefont {B.~J.}\ \bibnamefont {LeRoy}}, \bibinfo {author}
  {\bibfnamefont {A.~H.}\ \bibnamefont {MacDonald}}, \ and\ \bibinfo {author}
  {\bibfnamefont {E.}~\bibnamefont {Tutuc}},\ }\href@noop {} {\bibfield
  {journal} {\bibinfo  {journal} {Proceedings of the National Academy of
  Sciences}\ }\textbf {\bibinfo {volume} {114}},\ \bibinfo {pages} {3364}
  (\bibinfo {year} {2017})}\BibitemShut {NoStop}%
\bibitem [{SI()}]{SI}%
  \BibitemOpen
  \href@noop {} {\ }\bibinfo {note} {See the supplementary
  materials.}\BibitemShut {Stop}%
\bibitem [{\citenamefont {Zhang}\ \emph {et~al.}(2009)\citenamefont {Zhang},
  \citenamefont {Tang}, \citenamefont {Girit}, \citenamefont {Hao},
  \citenamefont {Martin}, \citenamefont {Zettl}, \citenamefont {Crommie},
  \citenamefont {Shen},\ and\ \citenamefont {Wang}}]{Zhang2009}%
  \BibitemOpen
  \bibfield  {author} {\bibinfo {author} {\bibfnamefont {Y.}~\bibnamefont
  {Zhang}}, \bibinfo {author} {\bibfnamefont {T.-T.}\ \bibnamefont {Tang}},
  \bibinfo {author} {\bibfnamefont {C.}~\bibnamefont {Girit}}, \bibinfo
  {author} {\bibfnamefont {Z.}~\bibnamefont {Hao}}, \bibinfo {author}
  {\bibfnamefont {M.~C.}\ \bibnamefont {Martin}}, \bibinfo {author}
  {\bibfnamefont {A.}~\bibnamefont {Zettl}}, \bibinfo {author} {\bibfnamefont
  {M.~F.}\ \bibnamefont {Crommie}}, \bibinfo {author} {\bibfnamefont {Y.~R.}\
  \bibnamefont {Shen}}, \ and\ \bibinfo {author} {\bibfnamefont
  {F.}~\bibnamefont {Wang}},\ }\href@noop {} {\bibfield  {journal} {\bibinfo
  {journal} {Nature}\ }\textbf {\bibinfo {volume} {459}},\ \bibinfo {pages}
  {820} (\bibinfo {year} {2009})}\BibitemShut {NoStop}%
\bibitem [{\citenamefont {Yoo}\ \emph {et~al.}(2018)\citenamefont {Yoo},
  \citenamefont {Zhang}, \citenamefont {Engelke}, \citenamefont {Cazeaux},
  \citenamefont {Sung}, \citenamefont {Hovden}, \citenamefont {Tsen},
  \citenamefont {Taniguchi}, \citenamefont {Watanabe}, \citenamefont {Yi},
  \citenamefont {Kim}, \citenamefont {Luskin}, \citenamefont {Tadmor},\ and\
  \citenamefont {Kim}}]{Yoo2018}%
  \BibitemOpen
  \bibfield  {author} {\bibinfo {author} {\bibfnamefont {H.}~\bibnamefont
  {Yoo}}, \bibinfo {author} {\bibfnamefont {K.}~\bibnamefont {Zhang}}, \bibinfo
  {author} {\bibfnamefont {R.}~\bibnamefont {Engelke}}, \bibinfo {author}
  {\bibfnamefont {P.}~\bibnamefont {Cazeaux}}, \bibinfo {author} {\bibfnamefont
  {S.~H.}\ \bibnamefont {Sung}}, \bibinfo {author} {\bibfnamefont
  {R.}~\bibnamefont {Hovden}}, \bibinfo {author} {\bibfnamefont {A.~W.}\
  \bibnamefont {Tsen}}, \bibinfo {author} {\bibfnamefont {T.}~\bibnamefont
  {Taniguchi}}, \bibinfo {author} {\bibfnamefont {K.}~\bibnamefont {Watanabe}},
  \bibinfo {author} {\bibfnamefont {G.-C.}\ \bibnamefont {Yi}}, \bibinfo
  {author} {\bibfnamefont {M.}~\bibnamefont {Kim}}, \bibinfo {author}
  {\bibfnamefont {M.}~\bibnamefont {Luskin}}, \bibinfo {author} {\bibfnamefont
  {E.~B.}\ \bibnamefont {Tadmor}}, \ and\ \bibinfo {author} {\bibfnamefont
  {P.}~\bibnamefont {Kim}},\ }\href@noop {} {\bibfield  {journal} {\bibinfo
  {journal} {arXiv:1804.03806}\ } (\bibinfo {year} {2018})}\BibitemShut
  {NoStop}%
\bibitem [{\citenamefont {Ryazanov}\ \emph {et~al.}(2001)\citenamefont
  {Ryazanov}, \citenamefont {Oboznov}, \citenamefont {Rusanov}, \citenamefont
  {Veretennikov}, \citenamefont {Golubov},\ and\ \citenamefont
  {Aarts}}]{Ryazanov2001}%
  \BibitemOpen
  \bibfield  {author} {\bibinfo {author} {\bibfnamefont {V.~V.}\ \bibnamefont
  {Ryazanov}}, \bibinfo {author} {\bibfnamefont {V.~A.}\ \bibnamefont
  {Oboznov}}, \bibinfo {author} {\bibfnamefont {A.~Y.}\ \bibnamefont
  {Rusanov}}, \bibinfo {author} {\bibfnamefont {A.~V.}\ \bibnamefont
  {Veretennikov}}, \bibinfo {author} {\bibfnamefont {A.~A.}\ \bibnamefont
  {Golubov}}, \ and\ \bibinfo {author} {\bibfnamefont {J.}~\bibnamefont
  {Aarts}},\ }\href@noop {} {\bibfield  {journal} {\bibinfo  {journal}
  {Physical Review Letters}\ }\textbf {\bibinfo {volume} {86}},\ \bibinfo
  {pages} {2427} (\bibinfo {year} {2001})}\BibitemShut {NoStop}%
\bibitem [{\citenamefont {Tsuei}\ and\ \citenamefont
  {Kirtley}(2000)}]{Tsuei2000}%
  \BibitemOpen
  \bibfield  {author} {\bibinfo {author} {\bibfnamefont {C.~C.}\ \bibnamefont
  {Tsuei}}\ and\ \bibinfo {author} {\bibfnamefont {J.~R.}\ \bibnamefont
  {Kirtley}},\ }\href@noop {} {\bibfield  {journal} {\bibinfo  {journal}
  {Reviews of Modern Physics}\ }\textbf {\bibinfo {volume} {72}},\ \bibinfo
  {pages} {969} (\bibinfo {year} {2000})}\BibitemShut {NoStop}%
\bibitem [{\citenamefont {Ribeiro-Palau}\ \emph {et~al.}(2018)\citenamefont
  {Ribeiro-Palau}, \citenamefont {Zhang}, \citenamefont {Watanbe},
  \citenamefont {Taniguchi}, \citenamefont {Hone},\ and\ \citenamefont
  {Dean}}]{Ribeiro2018}%
  \BibitemOpen
  \bibfield  {author} {\bibinfo {author} {\bibfnamefont {R.}~\bibnamefont
  {Ribeiro-Palau}}, \bibinfo {author} {\bibfnamefont {C.}~\bibnamefont
  {Zhang}}, \bibinfo {author} {\bibfnamefont {K.}~\bibnamefont {Watanbe}},
  \bibinfo {author} {\bibfnamefont {T.}~\bibnamefont {Taniguchi}}, \bibinfo
  {author} {\bibfnamefont {J.}~\bibnamefont {Hone}}, \ and\ \bibinfo {author}
  {\bibfnamefont {C.~R.}\ \bibnamefont {Dean}},\ }\href@noop {} {\bibfield
  {journal} {\bibinfo  {journal} {Science}\ }\textbf {\bibinfo {volume}
  {361}},\ \bibinfo {pages} {690} (\bibinfo {year} {2018})}\BibitemShut
  {NoStop}%
\bibitem [{\citenamefont {Yuan}\ and\ \citenamefont {Fu}(2018)}]{Yuan2018}%
  \BibitemOpen
  \bibfield  {author} {\bibinfo {author} {\bibfnamefont {N.~F.~Q.}\
  \bibnamefont {Yuan}}\ and\ \bibinfo {author} {\bibfnamefont {L.}~\bibnamefont
  {Fu}},\ }\href@noop {} {\bibfield  {journal} {\bibinfo  {journal} {Physical
  Review B}\ }\textbf {\bibinfo {volume} {98}},\ \bibinfo {pages} {045103}
  (\bibinfo {year} {2018})}\BibitemShut {NoStop}%
\bibitem [{\citenamefont {Song}\ \emph {et~al.}(2018)\citenamefont {Song},
  \citenamefont {Wang}, \citenamefont {Shi}, \citenamefont {Li}, \citenamefont
  {Fang},\ and\ \citenamefont {Bernevig}}]{Song2018}%
  \BibitemOpen
  \bibfield  {author} {\bibinfo {author} {\bibfnamefont {Z.}~\bibnamefont
  {Song}}, \bibinfo {author} {\bibfnamefont {Z.}~\bibnamefont {Wang}}, \bibinfo
  {author} {\bibfnamefont {W.}~\bibnamefont {Shi}}, \bibinfo {author}
  {\bibfnamefont {G.}~\bibnamefont {Li}}, \bibinfo {author} {\bibfnamefont
  {C.}~\bibnamefont {Fang}}, \ and\ \bibinfo {author} {\bibfnamefont {B.~A.}\
  \bibnamefont {Bernevig}},\ }\href@noop {} {\bibfield  {journal} {\bibinfo
  {journal} {arXiv:1807.10676}\ } (\bibinfo {year} {2018})}\BibitemShut
  {NoStop}%
\bibitem [{\citenamefont {Hejazi}\ \emph {et~al.}(2018)\citenamefont {Hejazi},
  \citenamefont {Liu}, \citenamefont {Shapourian}, \citenamefont {Chen},\ and\
  \citenamefont {Balents}}]{Hejazi2018}%
  \BibitemOpen
  \bibfield  {author} {\bibinfo {author} {\bibfnamefont {K.}~\bibnamefont
  {Hejazi}}, \bibinfo {author} {\bibfnamefont {C.}~\bibnamefont {Liu}},
  \bibinfo {author} {\bibfnamefont {H.}~\bibnamefont {Shapourian}}, \bibinfo
  {author} {\bibfnamefont {X.}~\bibnamefont {Chen}}, \ and\ \bibinfo {author}
  {\bibfnamefont {L.}~\bibnamefont {Balents}},\ }\href@noop {} {\bibfield
  {journal} {\bibinfo  {journal} {arXiv:1808.01568}\ } (\bibinfo {year}
  {2018})}\BibitemShut {NoStop}%
\bibitem [{\citenamefont {Xu}\ and\ \citenamefont {Balents}(2018)}]{Xu2018}%
  \BibitemOpen
  \bibfield  {author} {\bibinfo {author} {\bibfnamefont {C.}~\bibnamefont
  {Xu}}\ and\ \bibinfo {author} {\bibfnamefont {L.}~\bibnamefont {Balents}},\
  }\href@noop {} {\bibfield  {journal} {\bibinfo  {journal} {arXiv:1803.08057}\
  } (\bibinfo {year} {2018})}\BibitemShut {NoStop}%
\bibitem [{\citenamefont {Po}\ \emph {et~al.}(2018{\natexlab{a}})\citenamefont
  {Po}, \citenamefont {Zou}, \citenamefont {Vishwanath},\ and\ \citenamefont
  {Senthil}}]{Po2018}%
  \BibitemOpen
  \bibfield  {author} {\bibinfo {author} {\bibfnamefont {H.~C.}\ \bibnamefont
  {Po}}, \bibinfo {author} {\bibfnamefont {L.}~\bibnamefont {Zou}}, \bibinfo
  {author} {\bibfnamefont {A.}~\bibnamefont {Vishwanath}}, \ and\ \bibinfo
  {author} {\bibfnamefont {T.}~\bibnamefont {Senthil}},\ }\href@noop {}
  {\bibfield  {journal} {\bibinfo  {journal} {arXiv:1803.09742}\ } (\bibinfo
  {year} {2018}{\natexlab{a}})}\BibitemShut {NoStop}%
\bibitem [{\citenamefont {Roy}\ and\ \citenamefont {Juricic}(2018)}]{Roy2018}%
  \BibitemOpen
  \bibfield  {author} {\bibinfo {author} {\bibfnamefont {B.}~\bibnamefont
  {Roy}}\ and\ \bibinfo {author} {\bibfnamefont {V.}~\bibnamefont {Juricic}},\
  }\href@noop {} {\bibfield  {journal} {\bibinfo  {journal} {arXiv:1803.11190}\
  } (\bibinfo {year} {2018})}\BibitemShut {NoStop}%
\bibitem [{\citenamefont {Baskaran}(2018)}]{Baskaran2018}%
  \BibitemOpen
  \bibfield  {author} {\bibinfo {author} {\bibfnamefont {G.}~\bibnamefont
  {Baskaran}},\ }\href@noop {} {\bibfield  {journal} {\bibinfo  {journal}
  {arXiv:1804.00627}\ } (\bibinfo {year} {2018})}\BibitemShut {NoStop}%
\bibitem [{\citenamefont {Isobe}\ \emph {et~al.}(2018)\citenamefont {Isobe},
  \citenamefont {Yuan},\ and\ \citenamefont {Fu}}]{Isobe2018}%
  \BibitemOpen
  \bibfield  {author} {\bibinfo {author} {\bibfnamefont {H.}~\bibnamefont
  {Isobe}}, \bibinfo {author} {\bibfnamefont {N.~F.~Q.}\ \bibnamefont {Yuan}},
  \ and\ \bibinfo {author} {\bibfnamefont {L.}~\bibnamefont {Fu}},\ }\href@noop
  {} {\bibfield  {journal} {\bibinfo  {journal} {arXiv:1805.06449}\ } (\bibinfo
  {year} {2018})}\BibitemShut {NoStop}%
\bibitem [{\citenamefont {Padhi}\ \emph {et~al.}(2018)\citenamefont {Padhi},
  \citenamefont {Setty},\ and\ \citenamefont {Phillips}}]{Padhi2018}%
  \BibitemOpen
  \bibfield  {author} {\bibinfo {author} {\bibfnamefont {B.}~\bibnamefont
  {Padhi}}, \bibinfo {author} {\bibfnamefont {C.}~\bibnamefont {Setty}}, \ and\
  \bibinfo {author} {\bibfnamefont {P.~W.}\ \bibnamefont {Phillips}},\
  }\href@noop {} {\bibfield  {journal} {\bibinfo  {journal} {arXiv:1804.01101}\
  } (\bibinfo {year} {2018})}\BibitemShut {NoStop}%
\bibitem [{\citenamefont {Ochi}\ \emph {et~al.}(2018)\citenamefont {Ochi},
  \citenamefont {Koshino},\ and\ \citenamefont {Kuroki}}]{Ochi2018}%
  \BibitemOpen
  \bibfield  {author} {\bibinfo {author} {\bibfnamefont {M.}~\bibnamefont
  {Ochi}}, \bibinfo {author} {\bibfnamefont {M.}~\bibnamefont {Koshino}}, \
  and\ \bibinfo {author} {\bibfnamefont {K.}~\bibnamefont {Kuroki}},\
  }\href@noop {} {\bibfield  {journal} {\bibinfo  {journal} {Physical Review
  B}\ }\textbf {\bibinfo {volume} {98}},\ \bibinfo {pages} {081102(R)}
  (\bibinfo {year} {2018})}\BibitemShut {NoStop}%
\bibitem [{\citenamefont {Guo}\ \emph {et~al.}(2018)\citenamefont {Guo},
  \citenamefont {Zhu}, \citenamefont {Feng},\ and\ \citenamefont
  {Scalettar}}]{Guo2018}%
  \BibitemOpen
  \bibfield  {author} {\bibinfo {author} {\bibfnamefont {H.}~\bibnamefont
  {Guo}}, \bibinfo {author} {\bibfnamefont {X.}~\bibnamefont {Zhu}}, \bibinfo
  {author} {\bibfnamefont {S.}~\bibnamefont {Feng}}, \ and\ \bibinfo {author}
  {\bibfnamefont {R.~T.}\ \bibnamefont {Scalettar}},\ }\href@noop {} {\bibfield
   {journal} {\bibinfo  {journal} {Physical Review B}\ }\textbf {\bibinfo
  {volume} {97}},\ \bibinfo {pages} {235453} (\bibinfo {year}
  {2018})}\BibitemShut {NoStop}%
\bibitem [{\citenamefont {Huang}\ \emph {et~al.}(2018)\citenamefont {Huang},
  \citenamefont {Zhang},\ and\ \citenamefont {Ma}}]{Huang2018}%
  \BibitemOpen
  \bibfield  {author} {\bibinfo {author} {\bibfnamefont {T.}~\bibnamefont
  {Huang}}, \bibinfo {author} {\bibfnamefont {L.}~\bibnamefont {Zhang}}, \ and\
  \bibinfo {author} {\bibfnamefont {T.}~\bibnamefont {Ma}},\ }\href@noop {}
  {\bibfield  {journal} {\bibinfo  {journal} {arXiv:1804.06096}\ } (\bibinfo
  {year} {2018})}\BibitemShut {NoStop}%
\bibitem [{\citenamefont {Zhang}(2018)}]{Zhang2018b}%
  \BibitemOpen
  \bibfield  {author} {\bibinfo {author} {\bibfnamefont {L.}~\bibnamefont
  {Zhang}},\ }\href@noop {} {\bibfield  {journal} {\bibinfo  {journal}
  {arXiv:1804.09047}\ } (\bibinfo {year} {2018})}\BibitemShut {NoStop}%
\bibitem [{\citenamefont {Ray}\ and\ \citenamefont {Das}(2018)}]{Ray2018}%
  \BibitemOpen
  \bibfield  {author} {\bibinfo {author} {\bibfnamefont {S.}~\bibnamefont
  {Ray}}\ and\ \bibinfo {author} {\bibfnamefont {T.}~\bibnamefont {Das}},\
  }\href@noop {} {\bibfield  {journal} {\bibinfo  {journal} {arXiv:1804.09674}\
  } (\bibinfo {year} {2018})}\BibitemShut {NoStop}%
\bibitem [{\citenamefont {Liu}\ \emph {et~al.}(2018)\citenamefont {Liu},
  \citenamefont {Zhang}, \citenamefont {Chen},\ and\ \citenamefont
  {Yang}}]{Liu2018}%
  \BibitemOpen
  \bibfield  {author} {\bibinfo {author} {\bibfnamefont {C.-C.}\ \bibnamefont
  {Liu}}, \bibinfo {author} {\bibfnamefont {L.-D.}\ \bibnamefont {Zhang}},
  \bibinfo {author} {\bibfnamefont {W.-Q.}\ \bibnamefont {Chen}}, \ and\
  \bibinfo {author} {\bibfnamefont {F.}~\bibnamefont {Yang}},\ }\href@noop {}
  {\bibfield  {journal} {\bibinfo  {journal} {arXiv:1804.10009}\ } (\bibinfo
  {year} {2018})}\BibitemShut {NoStop}%
\bibitem [{\citenamefont {Fidrysiak}\ \emph {et~al.}(2018)\citenamefont
  {Fidrysiak}, \citenamefont {Zegrodnik},\ and\ \citenamefont
  {Spa\l{}ek}}]{Fidrysiak2018}%
  \BibitemOpen
  \bibfield  {author} {\bibinfo {author} {\bibfnamefont {M.}~\bibnamefont
  {Fidrysiak}}, \bibinfo {author} {\bibfnamefont {M.}~\bibnamefont
  {Zegrodnik}}, \ and\ \bibinfo {author} {\bibfnamefont {J.}~\bibnamefont
  {Spa\l{}ek}},\ }\href@noop {} {\bibfield  {journal} {\bibinfo  {journal}
  {arXiv:1805.01179}\ } (\bibinfo {year} {2018})}\BibitemShut {NoStop}%
\bibitem [{\citenamefont {Rademaker}\ and\ \citenamefont
  {Mellado}(2018)}]{Rademaker2018}%
  \BibitemOpen
  \bibfield  {author} {\bibinfo {author} {\bibfnamefont {L.}~\bibnamefont
  {Rademaker}}\ and\ \bibinfo {author} {\bibfnamefont {P.}~\bibnamefont
  {Mellado}},\ }\href@noop {} {\bibfield  {journal} {\bibinfo  {journal}
  {arXiv:1805.05294}\ } (\bibinfo {year} {2018})}\BibitemShut {NoStop}%
\bibitem [{\citenamefont {Kennes}\ \emph {et~al.}(2018)\citenamefont {Kennes},
  \citenamefont {Lischner},\ and\ \citenamefont {Karrasch}}]{Kennes2018}%
  \BibitemOpen
  \bibfield  {author} {\bibinfo {author} {\bibfnamefont {D.~M.}\ \bibnamefont
  {Kennes}}, \bibinfo {author} {\bibfnamefont {J.}~\bibnamefont {Lischner}}, \
  and\ \bibinfo {author} {\bibfnamefont {C.}~\bibnamefont {Karrasch}},\
  }\href@noop {} {\bibfield  {journal} {\bibinfo  {journal} {arXiv:1805.06310}\
  } (\bibinfo {year} {2018})}\BibitemShut {NoStop}%
\bibitem [{\citenamefont {You}\ and\ \citenamefont
  {Vishwanath}(2018)}]{You2018}%
  \BibitemOpen
  \bibfield  {author} {\bibinfo {author} {\bibfnamefont {Y.}~\bibnamefont
  {You}}\ and\ \bibinfo {author} {\bibfnamefont {A.}~\bibnamefont
  {Vishwanath}},\ }\href@noop {} {\bibfield  {journal} {\bibinfo  {journal}
  {arXiv:1805.06867}\ } (\bibinfo {year} {2018})}\BibitemShut {NoStop}%
\bibitem [{\citenamefont {Gonz\'alez}\ and\ \citenamefont
  {Stauber}(2018)}]{Gonzalez2018}%
  \BibitemOpen
  \bibfield  {author} {\bibinfo {author} {\bibfnamefont {J.}~\bibnamefont
  {Gonz\'alez}}\ and\ \bibinfo {author} {\bibfnamefont {T.}~\bibnamefont
  {Stauber}},\ }\href@noop {} {\bibfield  {journal} {\bibinfo  {journal}
  {arXiv:1807.01275}\ } (\bibinfo {year} {2018})}\BibitemShut {NoStop}%
\bibitem [{\citenamefont {Su}\ and\ \citenamefont {Lin}(2018)}]{Su2018}%
  \BibitemOpen
  \bibfield  {author} {\bibinfo {author} {\bibfnamefont {Y.}~\bibnamefont
  {Su}}\ and\ \bibinfo {author} {\bibfnamefont {S.-Z.}\ \bibnamefont {Lin}},\
  }\href@noop {} {\bibfield  {journal} {\bibinfo  {journal} {arXiv:1807.02196}\
  } (\bibinfo {year} {2018})}\BibitemShut {NoStop}%
\bibitem [{\citenamefont {Sherkunov}\ and\ \citenamefont
  {Betouras}(2018)}]{Sherkunov2018}%
  \BibitemOpen
  \bibfield  {author} {\bibinfo {author} {\bibfnamefont {Y.}~\bibnamefont
  {Sherkunov}}\ and\ \bibinfo {author} {\bibfnamefont {J.~J.}\ \bibnamefont
  {Betouras}},\ }\href@noop {} {\bibfield  {journal} {\bibinfo  {journal}
  {arXiv:1807.05524}\ } (\bibinfo {year} {2018})}\BibitemShut {NoStop}%
\bibitem [{\citenamefont {Dodaro}\ \emph {et~al.}(2018)\citenamefont {Dodaro},
  \citenamefont {Kivelson}, \citenamefont {Schattner}, \citenamefont {Sun},\
  and\ \citenamefont {Wang}}]{Dodaro2018}%
  \BibitemOpen
  \bibfield  {author} {\bibinfo {author} {\bibfnamefont {J.~F.}\ \bibnamefont
  {Dodaro}}, \bibinfo {author} {\bibfnamefont {S.~A.}\ \bibnamefont
  {Kivelson}}, \bibinfo {author} {\bibfnamefont {Y.}~\bibnamefont {Schattner}},
  \bibinfo {author} {\bibfnamefont {X.-Q.}\ \bibnamefont {Sun}}, \ and\
  \bibinfo {author} {\bibfnamefont {C.}~\bibnamefont {Wang}},\ }\href@noop {}
  {\bibfield  {journal} {\bibinfo  {journal} {arXiv:1804.03162}\ } (\bibinfo
  {year} {2018})}\BibitemShut {NoStop}%
\bibitem [{\citenamefont {Peltonen}\ \emph {et~al.}(2018)\citenamefont
  {Peltonen}, \citenamefont {Ojaj\"{a}rvi},\ and\ \citenamefont
  {Heikkil\"{a}}}]{Peltonen2018}%
  \BibitemOpen
  \bibfield  {author} {\bibinfo {author} {\bibfnamefont {T.~J.}\ \bibnamefont
  {Peltonen}}, \bibinfo {author} {\bibfnamefont {R.}~\bibnamefont
  {Ojaj\"{a}rvi}}, \ and\ \bibinfo {author} {\bibfnamefont {T.~T.}\
  \bibnamefont {Heikkil\"{a}}},\ }\href@noop {} {\bibfield  {journal} {\bibinfo
   {journal} {arXiv:1805.01039}\ } (\bibinfo {year} {2018})}\BibitemShut
  {NoStop}%
\bibitem [{\citenamefont {Wu}\ \emph {et~al.}(2018{\natexlab{b}})\citenamefont
  {Wu}, \citenamefont {MacDonald},\ and\ \citenamefont {Martin}}]{Wu2018b}%
  \BibitemOpen
  \bibfield  {author} {\bibinfo {author} {\bibfnamefont {F.}~\bibnamefont
  {Wu}}, \bibinfo {author} {\bibfnamefont {A.~H.}\ \bibnamefont {MacDonald}}, \
  and\ \bibinfo {author} {\bibfnamefont {I.}~\bibnamefont {Martin}},\
  }\href@noop {} {\bibfield  {journal} {\bibinfo  {journal} {arXiv:1805.08735}\
  } (\bibinfo {year} {2018}{\natexlab{b}})}\BibitemShut {NoStop}%
\bibitem [{\citenamefont {Lian}\ \emph {et~al.}(2018)\citenamefont {Lian},
  \citenamefont {Wang},\ and\ \citenamefont {Andrei~Bernevig}}]{Lian2018}%
  \BibitemOpen
  \bibfield  {author} {\bibinfo {author} {\bibfnamefont {B.}~\bibnamefont
  {Lian}}, \bibinfo {author} {\bibfnamefont {Z.}~\bibnamefont {Wang}}, \ and\
  \bibinfo {author} {\bibfnamefont {B.}~\bibnamefont {Andrei~Bernevig}},\
  }\href@noop {} {\bibfield  {journal} {\bibinfo  {journal} {arXiv:1807.04382}\
  } (\bibinfo {year} {2018})}\BibitemShut {NoStop}%
\bibitem [{\citenamefont {Laksono}\ \emph {et~al.}(2017)\citenamefont
  {Laksono}, \citenamefont {Leaw}, \citenamefont {Reaves}, \citenamefont
  {Singh}, \citenamefont {Wang}, \citenamefont {Adam},\ and\ \citenamefont
  {Gu}}]{Laksono2018}%
  \BibitemOpen
  \bibfield  {author} {\bibinfo {author} {\bibfnamefont {E.}~\bibnamefont
  {Laksono}}, \bibinfo {author} {\bibfnamefont {J.~N.}\ \bibnamefont {Leaw}},
  \bibinfo {author} {\bibfnamefont {A.}~\bibnamefont {Reaves}}, \bibinfo
  {author} {\bibfnamefont {M.}~\bibnamefont {Singh}}, \bibinfo {author}
  {\bibfnamefont {X.}~\bibnamefont {Wang}}, \bibinfo {author} {\bibfnamefont
  {S.}~\bibnamefont {Adam}}, \ and\ \bibinfo {author} {\bibfnamefont
  {X.}~\bibnamefont {Gu}},\ }\href@noop {} {\bibfield  {journal} {\bibinfo
  {journal} {Solid State Communications}\ }\textbf {\bibinfo {volume} {282}},\
  \bibinfo {pages} {38} (\bibinfo {year} {2017})}\BibitemShut {NoStop}%
\bibitem [{\citenamefont {Wang}\ \emph {et~al.}(2013)\citenamefont {Wang},
  \citenamefont {Meric}, \citenamefont {Huang}, \citenamefont {Gao},
  \citenamefont {Gao}, \citenamefont {Tran}, \citenamefont {Taniguchi},
  \citenamefont {Watanabe}, \citenamefont {Campos}, \citenamefont {Muller},
  \citenamefont {Guo}, \citenamefont {Kim}, \citenamefont {Hone}, \citenamefont
  {Shepard},\ and\ \citenamefont {Dean}}]{Wang2013}%
  \BibitemOpen
  \bibfield  {author} {\bibinfo {author} {\bibfnamefont {L.}~\bibnamefont
  {Wang}}, \bibinfo {author} {\bibfnamefont {I.}~\bibnamefont {Meric}},
  \bibinfo {author} {\bibfnamefont {P.~Y.}\ \bibnamefont {Huang}}, \bibinfo
  {author} {\bibfnamefont {Q.}~\bibnamefont {Gao}}, \bibinfo {author}
  {\bibfnamefont {Y.}~\bibnamefont {Gao}}, \bibinfo {author} {\bibfnamefont
  {H.}~\bibnamefont {Tran}}, \bibinfo {author} {\bibfnamefont {T.}~\bibnamefont
  {Taniguchi}}, \bibinfo {author} {\bibfnamefont {K.}~\bibnamefont {Watanabe}},
  \bibinfo {author} {\bibfnamefont {L.~M.}\ \bibnamefont {Campos}}, \bibinfo
  {author} {\bibfnamefont {D.~A.}\ \bibnamefont {Muller}}, \bibinfo {author}
  {\bibfnamefont {J.}~\bibnamefont {Guo}}, \bibinfo {author} {\bibfnamefont
  {P.}~\bibnamefont {Kim}}, \bibinfo {author} {\bibfnamefont {J.}~\bibnamefont
  {Hone}}, \bibinfo {author} {\bibfnamefont {K.~L.}\ \bibnamefont {Shepard}}, \
  and\ \bibinfo {author} {\bibfnamefont {C.~R.}\ \bibnamefont {Dean}},\
  }\href@noop {} {\bibfield  {journal} {\bibinfo  {journal} {Science}\ }\textbf
  {\bibinfo {volume} {342}},\ \bibinfo {pages} {614} (\bibinfo {year}
  {2013})}\BibitemShut {NoStop}%
\bibitem [{\citenamefont {Kang}\ and\ \citenamefont {Vafek}(2018)}]{Kang2018}%
  \BibitemOpen
  \bibfield  {author} {\bibinfo {author} {\bibfnamefont {J.}~\bibnamefont
  {Kang}}\ and\ \bibinfo {author} {\bibfnamefont {O.}~\bibnamefont {Vafek}},\
  }\href@noop {} {\bibfield  {journal} {\bibinfo  {journal} {arXiv:1805.04918}\
  } (\bibinfo {year} {2018})}\BibitemShut {NoStop}%
\bibitem [{\citenamefont {Koshino}\ \emph {et~al.}(2018)\citenamefont
  {Koshino}, \citenamefont {Yuan}, \citenamefont {Koretsune}, \citenamefont
  {Ochi}, \citenamefont {Kuroki},\ and\ \citenamefont {Fu}}]{Koshino2018}%
  \BibitemOpen
  \bibfield  {author} {\bibinfo {author} {\bibfnamefont {M.}~\bibnamefont
  {Koshino}}, \bibinfo {author} {\bibfnamefont {N.~F.~Q.}\ \bibnamefont
  {Yuan}}, \bibinfo {author} {\bibfnamefont {T.}~\bibnamefont {Koretsune}},
  \bibinfo {author} {\bibfnamefont {M.}~\bibnamefont {Ochi}}, \bibinfo {author}
  {\bibfnamefont {K.}~\bibnamefont {Kuroki}}, \ and\ \bibinfo {author}
  {\bibfnamefont {L.}~\bibnamefont {Fu}},\ }\href@noop {} {\bibfield  {journal}
  {\bibinfo  {journal} {arXiv:1805.06819}\ } (\bibinfo {year}
  {2018})}\BibitemShut {NoStop}%
\bibitem [{\citenamefont {Pal}(2018)}]{Pal2018}%
  \BibitemOpen
  \bibfield  {author} {\bibinfo {author} {\bibfnamefont {H.~K.}\ \bibnamefont
  {Pal}},\ }\href@noop {} {\bibfield  {journal} {\bibinfo  {journal}
  {arXiv:1805.08803}\ } (\bibinfo {year} {2018})}\BibitemShut {NoStop}%
\bibitem [{\citenamefont {Guinea}\ and\ \citenamefont
  {Walet}(2018)}]{Guinea2018}%
  \BibitemOpen
  \bibfield  {author} {\bibinfo {author} {\bibfnamefont {F.}~\bibnamefont
  {Guinea}}\ and\ \bibinfo {author} {\bibfnamefont {N.~R.}\ \bibnamefont
  {Walet}},\ }\href@noop {} {\bibfield  {journal} {\bibinfo  {journal}
  {arXiv:1806.05990}\ } (\bibinfo {year} {2018})}\BibitemShut {NoStop}%
\bibitem [{\citenamefont {Zou}\ \emph {et~al.}(2018)\citenamefont {Zou},
  \citenamefont {C.}, \citenamefont {Vishwanath},\ and\ \citenamefont
  {Senthil}}]{Zou2018}%
  \BibitemOpen
  \bibfield  {author} {\bibinfo {author} {\bibfnamefont {L.}~\bibnamefont
  {Zou}}, \bibinfo {author} {\bibfnamefont {P.~H.}\ \bibnamefont {C.}},
  \bibinfo {author} {\bibfnamefont {A.}~\bibnamefont {Vishwanath}}, \ and\
  \bibinfo {author} {\bibfnamefont {T.}~\bibnamefont {Senthil}},\ }\href@noop
  {} {\bibfield  {journal} {\bibinfo  {journal} {arXiv:1806.07873}\ } (\bibinfo
  {year} {2018})}\BibitemShut {NoStop}%
\bibitem [{\citenamefont {Po}\ \emph {et~al.}(2018{\natexlab{b}})\citenamefont
  {Po}, \citenamefont {Zou}, \citenamefont {Senthil},\ and\ \citenamefont
  {Vishwanath}}]{Po2018b}%
  \BibitemOpen
  \bibfield  {author} {\bibinfo {author} {\bibfnamefont {H.~C.}\ \bibnamefont
  {Po}}, \bibinfo {author} {\bibfnamefont {L.}~\bibnamefont {Zou}}, \bibinfo
  {author} {\bibfnamefont {T.}~\bibnamefont {Senthil}}, \ and\ \bibinfo
  {author} {\bibfnamefont {A.}~\bibnamefont {Vishwanath}},\ }\href@noop {}
  {\bibfield  {journal} {\bibinfo  {journal} {arXiv:1808.02482}\ } (\bibinfo
  {year} {2018}{\natexlab{b}})}\BibitemShut {NoStop}%
\bibitem [{\citenamefont {Tarnopolsky}\ \emph {et~al.}(2018)\citenamefont
  {Tarnopolsky}, \citenamefont {Kruchkov},\ and\ \citenamefont
  {Vishwanath}}]{Tarnopolsky2018}%
  \BibitemOpen
  \bibfield  {author} {\bibinfo {author} {\bibfnamefont {G.}~\bibnamefont
  {Tarnopolsky}}, \bibinfo {author} {\bibfnamefont {A.~J.}\ \bibnamefont
  {Kruchkov}}, \ and\ \bibinfo {author} {\bibfnamefont {A.}~\bibnamefont
  {Vishwanath}},\ }\href@noop {} {\bibfield  {journal} {\bibinfo  {journal}
  {arXiv:1808.05250}\ } (\bibinfo {year} {2018})}\BibitemShut {NoStop}%
\bibitem [{\citenamefont {Ahn}\ \emph {et~al.}(2018)\citenamefont {Ahn},
  \citenamefont {Park},\ and\ \citenamefont {Yang}}]{Ahn2018}%
  \BibitemOpen
  \bibfield  {author} {\bibinfo {author} {\bibfnamefont {J.}~\bibnamefont
  {Ahn}}, \bibinfo {author} {\bibfnamefont {S.}~\bibnamefont {Park}}, \ and\
  \bibinfo {author} {\bibfnamefont {B.-J.}\ \bibnamefont {Yang}},\ }\href@noop
  {} {\bibfield  {journal} {\bibinfo  {journal} {arXiv:1808.05375}\ } (\bibinfo
  {year} {2018})}\BibitemShut {NoStop}%
\bibitem [{\citenamefont {Xu}\ \emph {et~al.}(2018)\citenamefont {Xu},
  \citenamefont {Law},\ and\ \citenamefont {Lee}}]{Xu2018b}%
  \BibitemOpen
  \bibfield  {author} {\bibinfo {author} {\bibfnamefont {X.~Y.}\ \bibnamefont
  {Xu}}, \bibinfo {author} {\bibfnamefont {K.~T.}\ \bibnamefont {Law}}, \ and\
  \bibinfo {author} {\bibfnamefont {P.~A.}\ \bibnamefont {Lee}},\ }\href@noop
  {} {\bibfield  {journal} {\bibinfo  {journal} {arXiv:1805.00478}\ } (\bibinfo
  {year} {2018})}\BibitemShut {NoStop}%
\bibitem [{\citenamefont {Wu}\ \emph {et~al.}(2018{\natexlab{c}})\citenamefont
  {Wu}, \citenamefont {Pawlak}, \citenamefont {Jian},\ and\ \citenamefont
  {Xu}}]{Wu2018c}%
  \BibitemOpen
  \bibfield  {author} {\bibinfo {author} {\bibfnamefont {X.-C.}\ \bibnamefont
  {Wu}}, \bibinfo {author} {\bibfnamefont {K.~A.}\ \bibnamefont {Pawlak}},
  \bibinfo {author} {\bibfnamefont {C.-M.}\ \bibnamefont {Jian}}, \ and\
  \bibinfo {author} {\bibfnamefont {C.}~\bibnamefont {Xu}},\ }\href@noop {}
  {\bibfield  {journal} {\bibinfo  {journal} {arXiv:1805.06906}\ } (\bibinfo
  {year} {2018}{\natexlab{c}})}\BibitemShut {NoStop}%
\bibitem [{\citenamefont {Pizarro}\ \emph {et~al.}(2018)\citenamefont
  {Pizarro}, \citenamefont {Calder\'on},\ and\ \citenamefont
  {Bascones}}]{Pizarro2018}%
  \BibitemOpen
  \bibfield  {author} {\bibinfo {author} {\bibfnamefont {J.~M.}\ \bibnamefont
  {Pizarro}}, \bibinfo {author} {\bibfnamefont {M.~J.}\ \bibnamefont
  {Calder\'on}}, \ and\ \bibinfo {author} {\bibfnamefont {E.}~\bibnamefont
  {Bascones}},\ }\href@noop {} {\bibfield  {journal} {\bibinfo  {journal}
  {arXiv:1805.07303}\ } (\bibinfo {year} {2018})}\BibitemShut {NoStop}%
\bibitem [{\citenamefont {Thomson}\ \emph {et~al.}(2018)\citenamefont
  {Thomson}, \citenamefont {Chatterjee}, \citenamefont {Schdev},\ and\
  \citenamefont {Scheurer}}]{Thomson2018}%
  \BibitemOpen
  \bibfield  {author} {\bibinfo {author} {\bibfnamefont {A.}~\bibnamefont
  {Thomson}}, \bibinfo {author} {\bibfnamefont {S.}~\bibnamefont {Chatterjee}},
  \bibinfo {author} {\bibfnamefont {S.}~\bibnamefont {Schdev}}, \ and\ \bibinfo
  {author} {\bibfnamefont {M.~S.}\ \bibnamefont {Scheurer}},\ }\href@noop {}
  {\bibfield  {journal} {\bibinfo  {journal} {Physical Review B}\ }\textbf
  {\bibinfo {volume} {98}},\ \bibinfo {pages} {075109} (\bibinfo {year}
  {2018})}\BibitemShut {NoStop}%
\end{thebibliography}%

\clearpage


\renewcommand{\thefigure}{S\arabic{figure}}
\renewcommand{\thesubsection}{S\arabic{subsection}}
\renewcommand{\theequation}{S\arabic{equation}}
\renewcommand{\thetable}{S\arabic{table}}
\setcounter{figure}{0} 
\setcounter{equation}{0}

\section*{Supplementary Information}

\subsection*{Methods}

\begin{figure*}[ht]
\includegraphics[width=6.9 in]{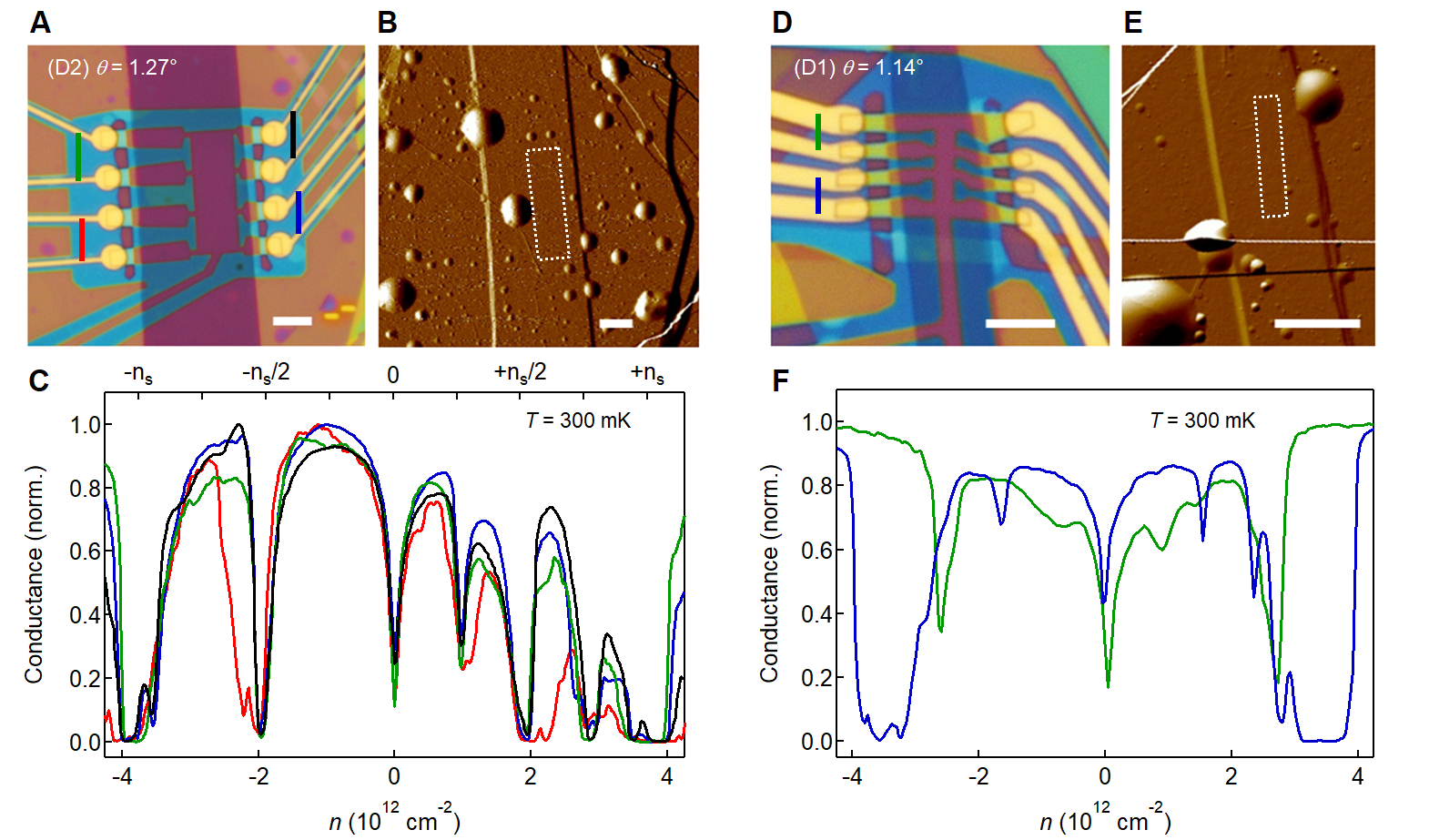} 
\caption{\textbf{Structural inhomogeneity in tBLG devices.}
(\textbf{A}) Optical microscope image of device D2.
(\textbf{B}) Atomic force microscope image of device D2 before etching. The outline of the Hall bar is shown in the white dashed line.
(\textbf{C}) Two-terminal conductance measured between the leads shown color-coded in (A) acquired at 2.21 GPa and $T$ = 300 mK. The density corresponding to full-filling $n_s$ is consistent across all contacts, although there is some apparent disorder in the bottom left contact in the device (red curve).
(\textbf{D}) Optical microscope and 
(\textbf{E}) AFM image of device D1. The outline of the Hall bar is shown in the white dashed line.
(\textbf{F}) Two-terminal conductance measurements, showing significant difference in the density needed to achieve full-filling between different sets of contacts. All scale bars are 5 $\mu$m.
}
\label{fig:S1}
\end{figure*}

All devices in this study consist of twisted bilayer graphene encapsulated between flakes of BN, typically between 30-60 nm thick. The entire structure is encapsulated between flakes of graphite which act as top and bottom gates, and rests on a Si/SiO$_2$ wafer. The heterostructure is assembled using standard dry-transfer techniques with a poly-propylene carbonate (PPC) film on a polydimethyl siloxane (PDMS) stamp~\cite{Wang2013}. The temperature is kept below 180 $^{\circ}$C throughout processing to prevent relaxation of the tBLG to the Bernal stacking configuration. Regions of the tBLG extend beyond both the top and bottom graphite gates and are set to a high density with a voltage on the Si gate (typically 5-40 V for SiO$_2$ thickness of $\sim$285 nm) to act as transparent contacts to the channel. Fig.~\ref{fig:S1}A shows an optical microscope image of device D2, and is qualitatively similar to all devices used in this study. Four-terminal resistance measurements are acquired with ac current excitation of 0.5-10 nA, and two-terminal conductance measurements are acquired with ac voltage excitations of 50-100 $\mu$V, using standard lock-in technique at 17.7Hz. Pressure measurements are performed following the procedures detailed in Ref.~\onlinecite{Yankowitz2018}. 

Table~\ref{table:table1} lists details of the three devices investigated in this study. To identify the twist angle in each device, we first estimate $n_s$ from the high-density insulating features. However, as these features are quite wide in gate voltage we then refine our estimate by measuring the density to which the sequence of quantum oscillations at half-filling project at $B$ = 0 (i.e. we extract $n_s/2$). We then determine the twist angle $\theta$ using the relation $n_s \approx \frac{8 \theta^2}{\sqrt{3} a^2}$, where $a$ = 0.246 nm is the lattice constant of graphene.

The tBLG is fabricated using the ``tear-and-stack'' method~\cite{Kim2016}, with a rotation of the transfer stage of $\sim$1.3$^{\circ}$. We find although the tBLG has a tendency to relax to the Bernal stacking configuration, flakes can slip during the transfer procedure and also result in twist angles larger than intended. Contamination between graphene layers which accrues during the transfer procedure aggregates into bubbles which are pushed out of the channel area during lamination of the heterostructure onto the Si/SiO$_2$ wafer, leaving a pristine atomic interface (see Fig.~\ref{fig:S1}B for an atomic force microscope image of device D2 before processing). We have found that the BLG twist angle varies from 0$^{\circ}$ to 2.25$^{\circ}$ although the stage is rotated by roughly the same amount during every transfer. We speculate there are three likely sources of this inconsistency. First, the top graphene layer may rotate on the top BN, as the friction between these crystals is small. Second, the bottom graphene layer has a tendency to ``jump'' off of the Si/SiO$_2$ wafer and onto the stack, during which it may rotate slightly. Third, the graphene may strain during the transfer procedure, resulting in a different effective area of the moir\'e unit cell for a given angle.

Although our device exhibits signatures of low charge disorder of  order 10$^{10}$ cm$^{-2}$, we find there are still markers of inhomogeneity in our devices. Fig.~\ref{fig:S1}C shows the two-terminal conductance measurement of device D2 between various neighboring pairs of contacts (color coded corresponding to the connecting lines in Fig.~\ref{fig:S1}A). While this device exhibits consistent moir\'e unit cell size between all contacts, as evidenced by the good alignment of insulating states, the bottom left contact also appears to have a region of slightly different moir\'e unit cell area, judging by the larger width of the insulating features in the red trace. The slightly inhomogeneous contact is left floating during all measurements presented in the main text, and we do not observe signatures of significant disorder elsewhere in the device. 

In contrast, device D1 shows more obvious signatures of structural disorder. Figs.~\ref{fig:S1}D and E show optical and atomic force microscope (AFM) images of the device, while Fig.~\ref{fig:S1}F shows a comparison of the conductance between neighboring contacts on the left-hand side of the device. In this device, we find a more obvious density offset of the full-filling insulating states, even though there are no obvious corresponding structural defects in the AFM image. Recent TEM imaging of small-angle tBLG shows spatial fluctuations in the moir\'e period over length scales smaller than 1 $\mu$m~\cite{Yoo2018} likely arising due to strain, consistent with our observation of the varying area of the moir\'e unit cell. For the measurements of device D1 presented in the main text, we choose four neighboring contacts with nearly indistinguishable two-terminal transport characteristics, however we are not able to detect strains on length scales smaller than the contact separation. Further, there may also be spatially inhomogeneous strains arising at the interfaces of the encapsulating BN layers.

\begin{center}
\begin{table*}
\centering
\caption{
Experimental details of the four devices in this study.}
\label{table:table1}
\vspace{7pt}
\begin{tabular}{ | C{40pt} | C{40pt} | C{80pt} | C{115pt} | C{105pt} | C{105pt} N |}
\hline
\textbf{Device} & \textbf{Angle} & \textbf{Pressure (GPa)} & \textbf{Base Temperature (mK)} & \textbf{Correlated Insulators} & \textbf{Superconductivity} & \\[10pt]
\hline
D1 & ~$\sim$1.14$^{\circ}$ & 0 & 10 & resistive states at $\pm n_s/2$ and $+3n_s/4$ & $T_c \approx$ 0.4 K for holes and $T_c \approx$ 0.25 K for electrons & \\
\hline
D2 & ~$\sim$1.27$^{\circ}$ & 0 & 300 & very weakly developing at $\pm n_s/2$ and $\pm 3n_s/4$ & none observed & \\
\hline
D2 & ~$\sim$1.27$^{\circ}$ & 1.33 & 300 & well-developed at $\pm n_s/2$ and $+ 3n_s/4$, resistive state at $+n_s/4$ & $T_c \approx$ 3.1 K for holes, still developing for electrons & \\
\hline
D2 & ~$\sim$1.27$^{\circ}$ & 2.21 & 300 & well-developed at $\pm n_s/2$ and $+ 3n_s/4$, resistive state at $+n_s/4$ & $T_c \approx$ 1.7 K for holes, still developing for electrons & \\
\hline
D3 & ~$\sim$1.10$^{\circ}$ & 0 & 300 & well-developed at $+ n_s/2$, resistive states at $+n_s/4$, $-n_s/2$, and $\pm 3n_s/4$ & none observed (likely due to spatial inhomogeneity) & \\
\hline
D4 & ~$\sim$1.55$^{\circ}$ & 0 & 300 & none observed & none observed & \\
\hline
\end{tabular}
\end{table*}
\end{center}

\subsection*{Phase diagram of device D2}

\begin{figure*}[ht]
\includegraphics[width=4.6 in]{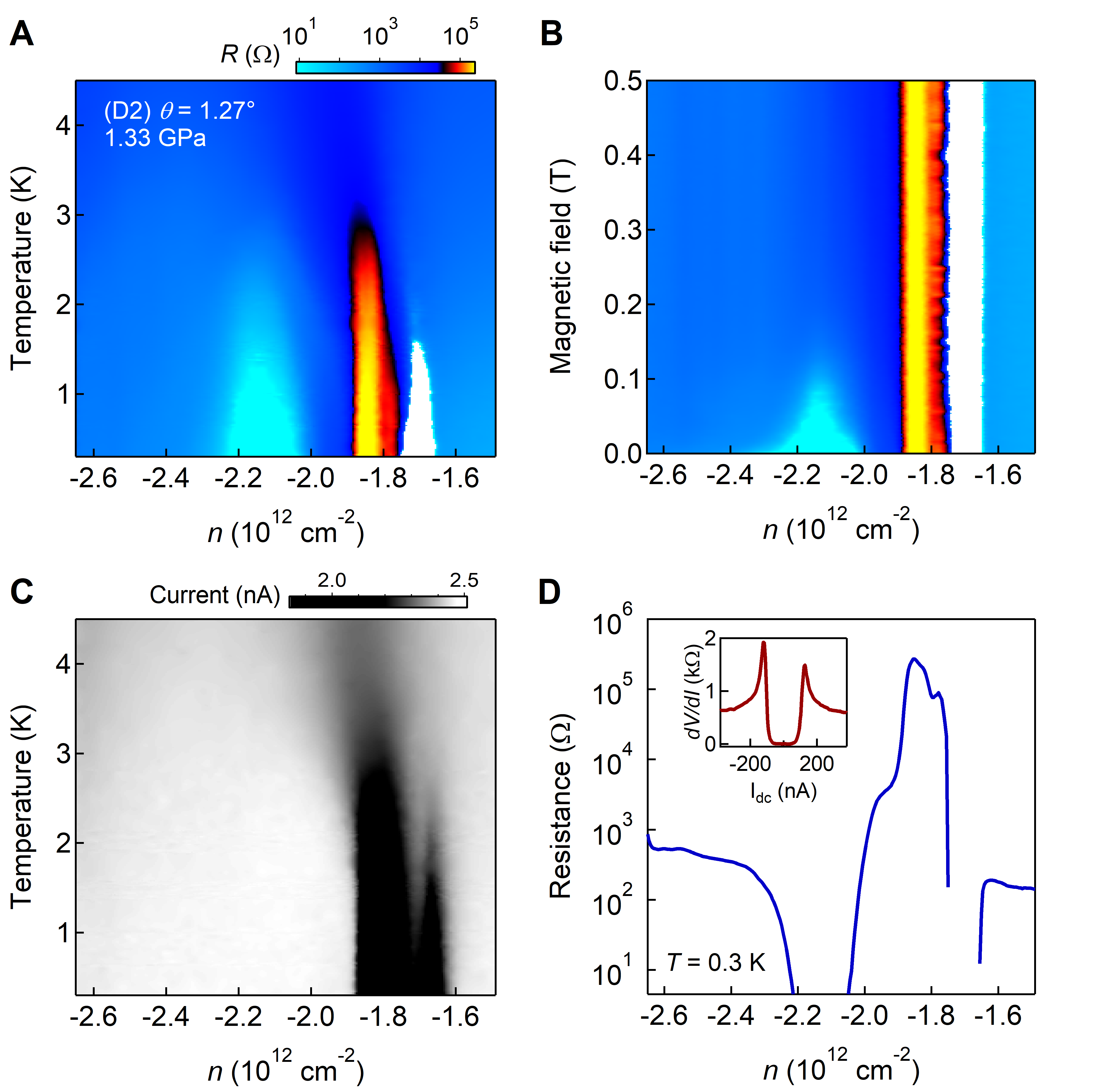} 
\caption{\textbf{Phase diagram of device D2 at 1.33 GPa.}
(\textbf{A}) Device resistance around $-n_s/2$ versus $T$ using contacts on the opposite side of the device from the comparable map in Fig.~\ref{fig:3}A of the main text. The region of artificially negative resistance is on the opposite side of the insulating state using these contacts, revealing more clearly an apparent metallic phase separating the superconductor and insulator. The apparent $T_c$ of the superconductor is lower with these contacts, although there is a weak region of lower resistance persisting up to the higher $T_c$ observed with the other set of contacts.
(\textbf{B}) Similar map versus $B$.
(\textbf{C}) Map of the current measured through the device simultaneous to the measurement acquired in (A). The measured current is attenuated (black regions) only when the device is insulating in panel (A), i.e. the red/yellow regions and the white region representing the artificially negative measured resistance.
(\textbf{D}) Resistance versus $n$ at $T$ = 300 mK and $B$ = 0.
(inset) $dV/dI$ as a function of $I_{dc}$ at $T$ = 300 mK and $B$ = 0.
}
\label{fig:S2}
\end{figure*}

Figs.~\ref{fig:S2}A and B show the resistance of device D2 around $-n_s/2$ as a function of temperature and magnetic field, respectively, at 1.33 GPa using a different set of contacts than those used to acquire Fig.~\ref{fig:3}A of the main text. In this case, there is a region of anomalously negative resistance (colored in white) at hole doping just less than $-n_s/2$. Fig.~\ref{fig:S2}C shows the corresponding measured current through the device on a saturated color scale. The black regions correspond to an attenuation of the current when the device becomes very resistive, and corresponds to the red/yellow and white regions of Fig.~\ref{fig:S2}A. The device appears to show slightly lower $T_c$ using these contacts, although there is a region of lower device resistance extending up to $\sim$3 K corresponding to the $T_c$ observed with the contact configuration used in Fig.~\ref{fig:3}A of the main text. Notably, the lack of anomalously negative resistance at hole doping just larger than $-n_s/2$ reveals more clearly an apparent metallic phase separating the superconductor and insulator. Fig.~\ref{fig:S2}D shows the resistance as a function of $n$ at $T$ = 300 mK. The inset shows the differential resistance $dV/dI$ as a function of $I_{dc}$ at $T$ = 300 mK and $B$ = 0.

\begin{figure*}[ht]
\includegraphics[width=4.6 in]{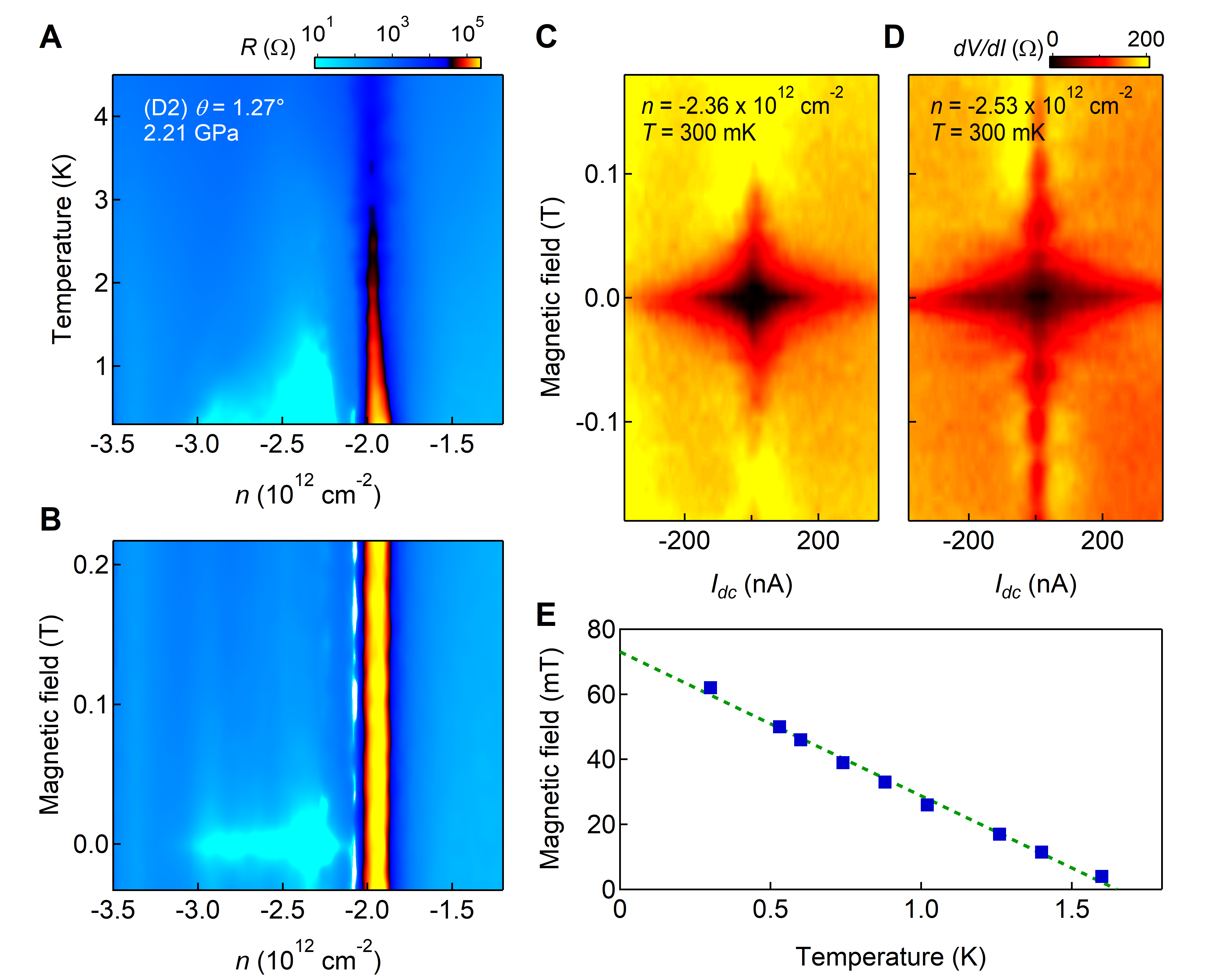} 
\caption{\textbf{Phase diagram of device D2 at 2.21 GPa.} 
(\textbf{A}) Device resistance around $-n_s/2$ versus $T$ with the same contact configuration used in Fig.~\ref{fig:3}A of the main text. The contact configuration used in Fig.~\ref{fig:S3} is qualitatively similar at this pressure.
(\textbf{B}) Similar map versus $B$.
(\textbf{C}) Differential resistance $dV/dI$ versus $I_{dc}$ and $B$ at $n = -2.36 \times 10^{12}$ cm$^{-2}$ and (\textbf{D}) $n = -2.53 \times 10^{12}$ cm$^{-2}$. $T$ = 300 mK for both (C) and (D). We observe very weak signatures of quantum interference at 2.21 GPa in (D) near the edge of the superconducting pocket.
(\textbf{E}) $H_{c2}$ versus temperature.
}
\label{fig:S3}
\end{figure*}

Figs.~\ref{fig:S3}A and B show similar measurements of the device resistance at 2.21 GPa, measured with the same contact configuration used in Figs.~\ref{fig:3}A and B in the main text (there is little difference between the two contact configurations at this pressure). We observe an asymmetry in the shape of the superconducting pocket with $n$, although it is not clear if this is an intrinsic property of the band structure at this pressure, or is instead arising due to a slight increase in device disorder. We observe the emergence of very weak quantum interference features in measurements of $dV/dI$ versus $I_{dc}$ and $B$ (Figs.~\ref{fig:S3}C and D), especially prominent in Fig.~\ref{fig:S3}D taken at $n = -2.6 \times 10^{12}$ cm$^{-2}$, near the minimum in the double-pocket structure of the superconductor. The device was removed from the pressure cell, cleaned with solvents, and reloaded between the measurements at 1.33 GPa and 2.21 GPa, and it is possible that this procedure resulting in slightly higher device disorder. However, it should be noted that quantum interference effects are weak an only arise near the edge of the superconducting pockets, therefore any influence of additional disorder should be small.

Fig.~\ref{fig:S3}E shows a measurement of $H_{c2}$ versus temperature under perpendicular magnetic field at 2.21 GPa, defined by the field at which we observe half the normal-state resistance. Similar to the observations in Ref.~\onlinecite{Cao2018b}, we find that $H_{c2}$ is linear in $T$. Following Ginzburg-Landau theory, we expect $H_{c2} = [\Phi_0/(2 \pi \xi^2_{GL}](1-T/T_c)$, where $\Phi_0 = h/(2e)$ is the superconducting flux quantum, $h$ is the Planck constant, and $\xi_{GL}$ is the Ginzburg-Landau superconducting coherence length, which we extract to be $\xi_{GL} \approx$ 68 nm at $T$ = 0. While we have not made a similar measurement at 1.33 GPa, we find an $H_{c2} \approx$ 0.21 mT at 300 mK, and assuming a similar linear decrease of $H_{c2}$ up to the critical temperature of approximately 3.15 K, we expect $\xi_{GL} \approx$ 38 nm at $T$ = 0. This is notably only a factor of a few larger than the moir\'e length scale itself, which is approximately 11 nm for device D2.   

\begin{figure*}[ht]
\includegraphics[width=4.6 in]{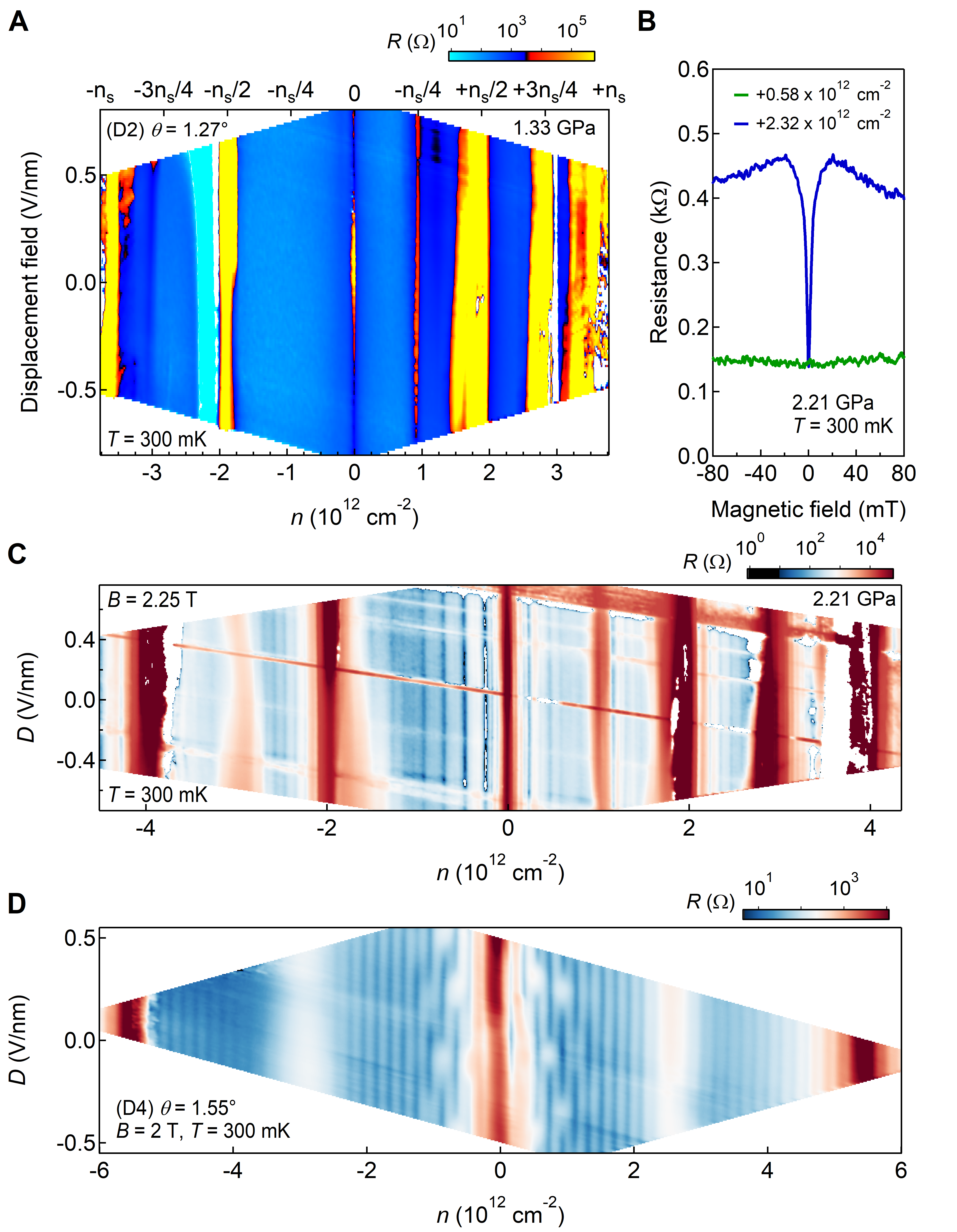} 
\caption{\textbf{Displacement field dependence of device D2 and comparison to device D4 with larger twist angle.}
(\textbf{A}) Resistance of device D2 as a function of $n$ and displacement field at 1.33 GPa and $B$ = 0.
(\textbf{B}) Resistance versus $B$ at optimal doping of the electron-type superconductor (blue curve) at 2.21 GPa and at low electron doping in the metallic phase (green curve) for comparison.
(\textbf{C}) $R_{xx}$ as a function of displacement field at $B$ = 2.25 T and 2.21 GPa. No Landau level transitions with $D$ are observed.
(\textbf{D}) $R_{xx}$ of device D4 as a as a function of displacement field at $B$ = 2 T and ambient pressure. Landau level transitions are observed around the CNP at various values of $D$. $T$ = 300 mK for all measurements.
}
\label{fig:S4}
\end{figure*}

Fig.~\ref{fig:S4}A shows a map of the device resistance as a function of $n$ and $D$ at $T$ = 300 mK and 1.33 GPa. In contrast to the comparable measurement of device D1 shown in Fig.~\ref{fig:1}D of the main text, we do not observe a transition in the state at $-n_s/2$ from an insulator to superconductor with $D$, and instead find a robust insulating state at all $D$. While some features are weakly tuned with $D$, in general the displacement field has little effect on the phase diagram of this device.

The blue curve in Fig.~\ref{fig:S4}B shows the device resistance as a function of $B$ for electron-doping slightly larger than $+n_s/2$ at $T$ = 300 mK and 2.21 GPa, showing a developing superconducting phase. In contrast, the device resistance is independent of $B$ for smaller electron doping (green curve) as anticipated for a metal.

Fig.~\ref{fig:S4}C shows a map of $R_{xx}$ at $B$ = 2.25 T and $T$ = 300 mK as a function of $n$ and $D$ at 2.21 GPa. We do not observe any Landau level crossings with displacement field. In contrast, devices with slightly larger twist angle exhibit numerous Landau level crossings with $D$. Fig.~\ref{fig:S4}D shows a comparable map acquired for device D4, with twist angle ~$\sim$1.55$^{\circ}$, acquired at $B$ = 2 T. We do not observe any signatures of correlated insulating states or superconductivity in this device at $B$ = 0. In a magnetic field, we observe eight-fold degeneracy of Landau levels near the CNP only at certain discrete values of $D$, and four-fold degeneracy for the remaining values of $D$. This suggests that the interlayer potential is able to lift the layer degeneracy of the Landau levels near the CNP, while at special values of $D$ they become coincident. We observe four-fold degeneracy of the Landau levels for higher values of $n$ with no transitions as a function of $D$. 

\subsection*{Quantum oscillations in device D2}

\begin{figure*}[p]
\includegraphics[width=6.9 in]{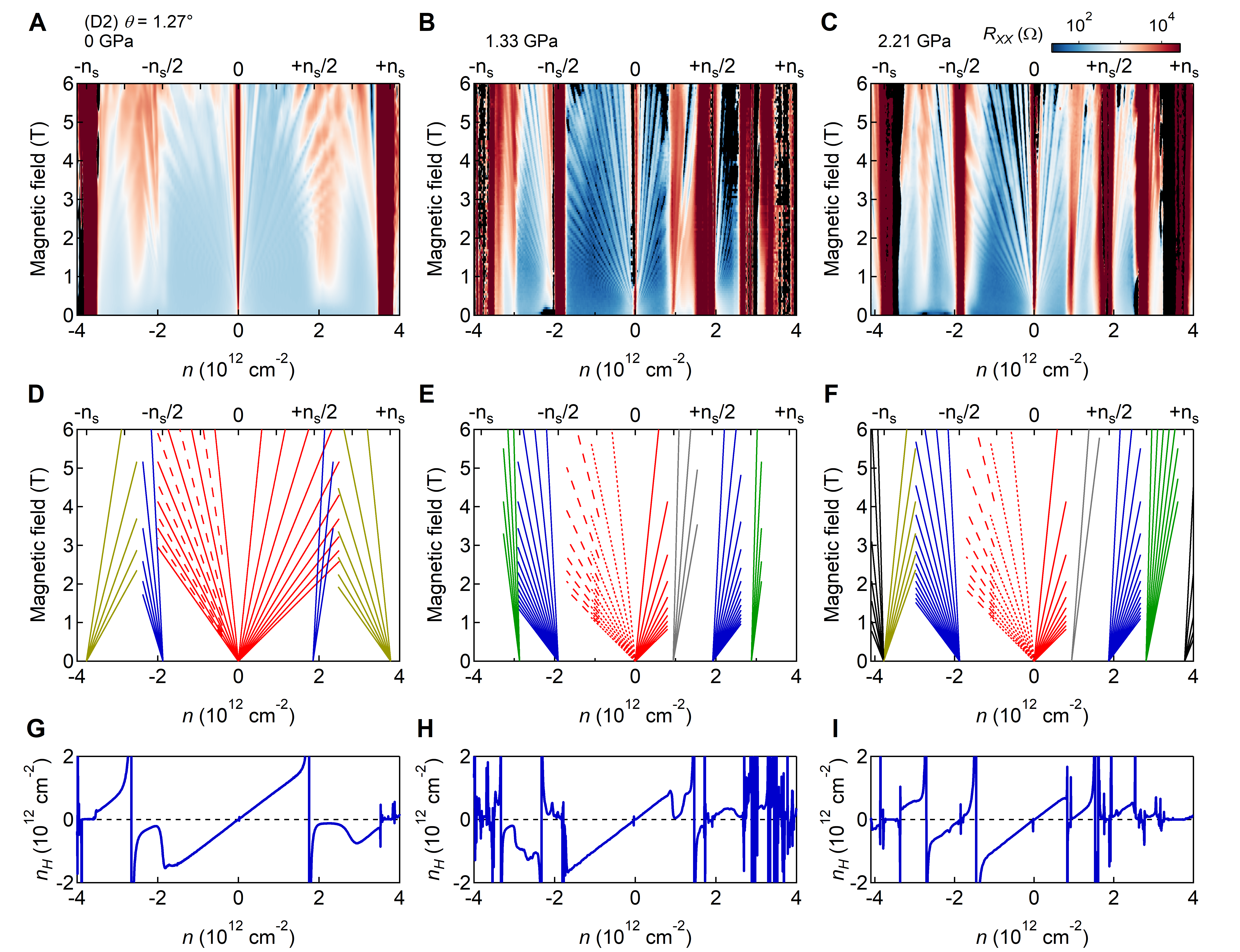} 
\caption{\textbf{Evolution of quantum oscillations in device D1 with pressure.} 
Landau fan diagram of device D2 at $T$ = 300 mK at 
(\textbf{A}) 0 GPa,
(\textbf{B}) 1.33 GPa, and
(\textbf{C}) 2.21 GPa.
In all cases, a sequence of quantum oscillations at low field emerge from the CNP with dominant degeneracy sequence of $\nu = \pm 4, \pm 8, \pm 12,...$, and a separate set of sequences emerge from $\pm n_s/2$ with dominant degeneracy sequence of $\nu = \pm 2, \pm 4, \pm 6,...$. Above $\pm 3n_s/4$, quantum oscillations may disperse either towards $\pm n_s$ or towards the CNP. In the former case, states at all integer filling factors are observed.
Schematic Landau level structures shown in 
(\textbf{D}) - (\textbf{F}) correspond to the fan diagrams in (\textbf{A}) - (\textbf{C}). Only the Landau levels surviving to the lowest fields are annotated.
(\textbf{G}) - (\textbf{I}) Hall density $n_H$ extracted from the low-field Hall resistance as a function of total charge density induced by the gate.
}
\label{fig:S5}
\end{figure*}

Figs.~\ref{fig:S5}A-C show the quantum oscillations in device D2 for 0 GPa, 1.33 GPa, and 2.21 GPa up to 6 T. Fig.~\ref{fig:S5}D-F show corresponding schematic versions of the observed Landau levels which persist to the lowest magnetic fields. We observe a dramatic change in the quantum oscillations between 0 GPa and 1.33 GPa, and more subtle differences between 1.33 GPa and 2.21 GPa. In all cases, quantum oscillations emerging from the CNP are colored in red and are four-fold degenerate. They all show an apparent shift in degeneracy as discussed in the main text around $-n_s/4$, although the transition region is rather wide and difficult to determine at 0 GPa. Quantum oscillations emerging from $+n_s/4$ are observed at 1.33 GPa and 2.21 GPa and are colored in gray. They are two-fold degenerate and odd dominant, although at 2.21 GPa the state at $\nu = +1$ is missing, and the degeneracy at high $\nu$ is difficult to determine. At $\pm n_s/2$ we observe sequences of two-fold, even-dominant quantum oscillations colored in blue. The sequence at $+n_s/2$ at 0 GPa is extremely weak, and the quantum oscillations from the CNP overlap as well. 

The quantum oscillations above $\pm 3n_s/4$ vary more substantially across measurements at different pressures. At 0 GPa, we observe sequences of four-fold degenerate quantum oscillations emerging from $\pm n_s$ and dispersing towards the CNP, colored in gold. We observe a similar sequence at $-n_s$ at 2.21 GPa, with an apparent two-fold degeneracy, although the latter is hard to judge because the quantum oscillations are obscured very near full filling. In contrast, at 1.33 GPa we observe sequences of quantum oscillations of similar strength at all integer filling factors emerging from $\pm 3n_s/4$, and at 2.21 GPa we observe a similar sequence emerging $+3n_s/4$. We also typically observe quantum oscillations developing above full filling, most clearly visible at 2.21 GPa and annotated by the black lines.

Figs.~\ref{fig:S5}G-I show the corresponding Hall density $n_H$ as a function of total carrier density induced by the gate $n$, measured from the slope of the Hall resistance $R_{xy}$ at low $B$ following $n = B/(eR_{xy})$. We find that while these two quantities match at low density, the magnitude and even sign of $n_H$ changes for higher filling, typically associated with the emergence of correlated insulating states in $R_{xx}$ and likely an indication of the formation of new, small Fermi surfaces at these densities.

Finally, we note that at all pressures the quantum oscillations from each respective Fermi surface project to nearly identical charge carrier densities. This demonstrates that pressure does not change the twist angle of the BLG, and instead only modifies the strength of the interlayer coupling, as has been observed previously in aligned graphene on BN heterostructures~\cite{Yankowitz2018}.

\subsection*{Temperature and current quenching of correlated states}

\begin{figure*}[ht]
\includegraphics[width=4.6 in]{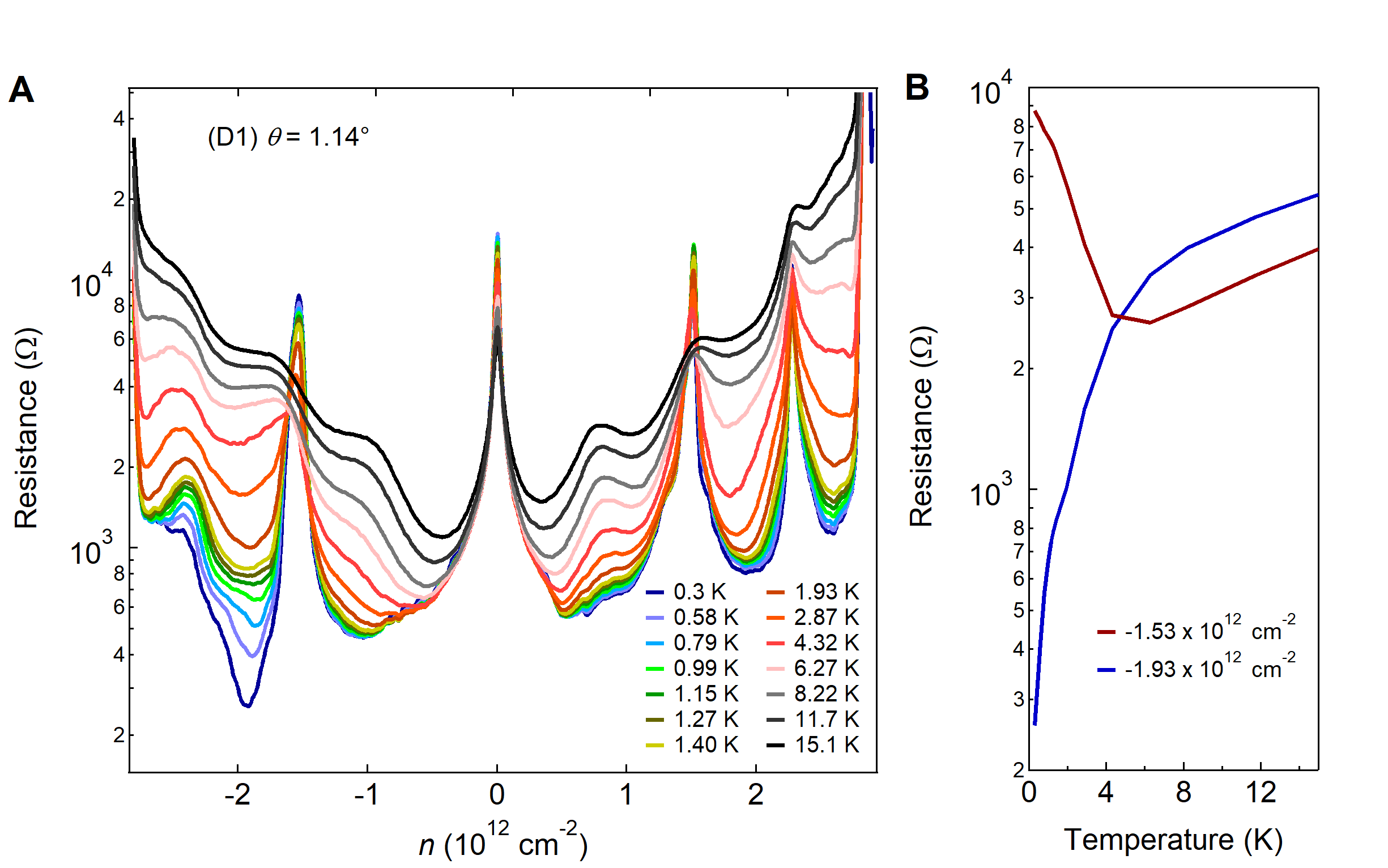} 
\caption{\textbf{Resistance as a function of temperature in device D1.}
(\textbf{A}) Resistance traces at various temperatures from 0.3 K to 15.1 K.
(\textbf{B}) Resistance versus temperature at $-n_s/2$ (red curve) and at optimal doping of the superconductor (blue curve).
}
\label{fig:S6}
\end{figure*}

\begin{figure*}[ht]
\includegraphics[width=3.6 in]{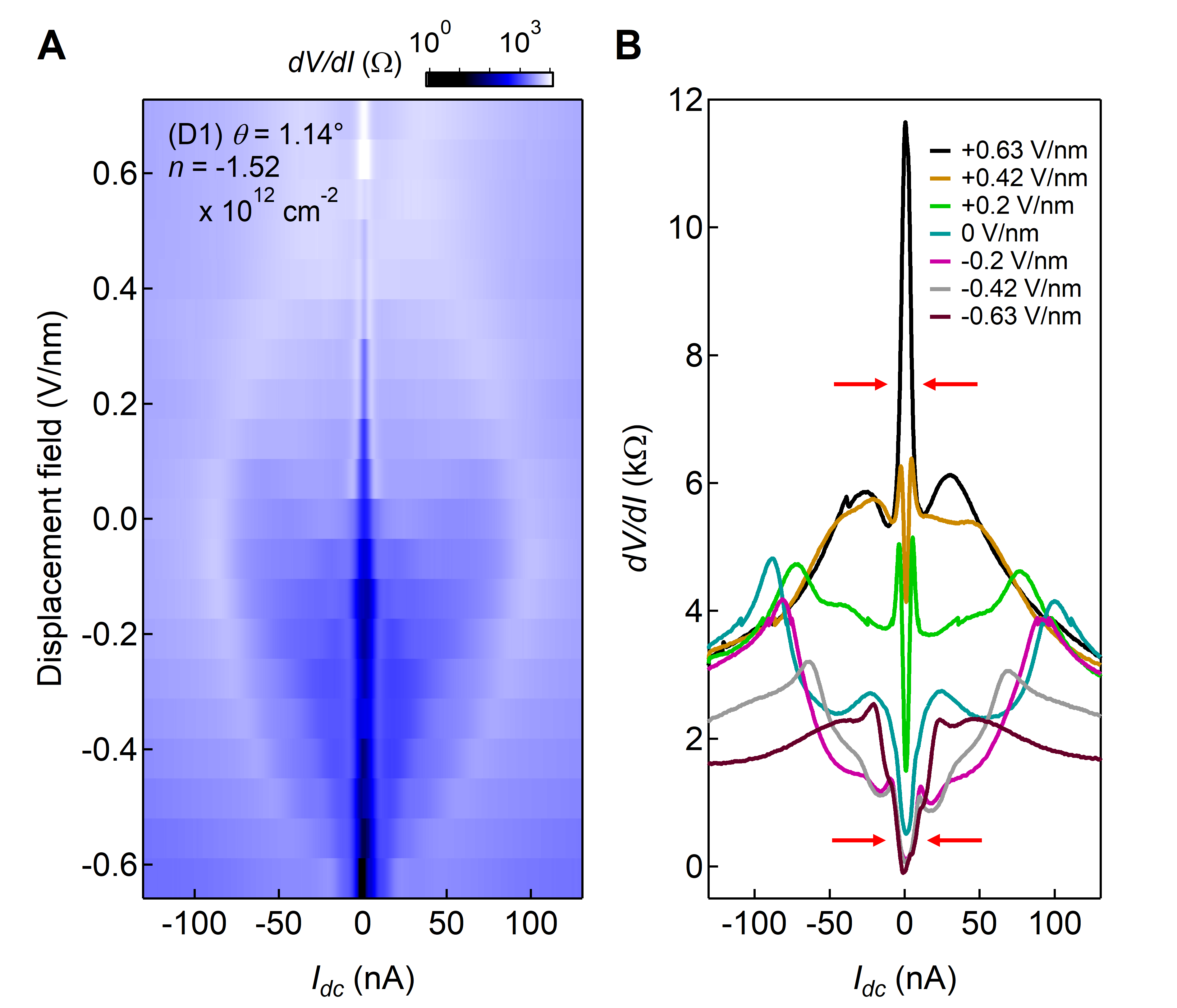} 
\caption{\textbf{Current breakdown of correlated phases in device D1.}
(\textbf{A}) $dV/dI$ as a function of $I_{dc}$ and $D$ at $n = -1.52 \times 10^{12}$ cm$^{-2}$, corresponding roughly to $-n_s/2$. The response evolves from an insulator at positive $D$ to a superconductor at negative $D$.
(\textbf{B}) Individual $dV/dI$ traces at various $D$. The superconducting and insulating states appear to be quenched at similar critical currents, annotated by the red arrows.
}
\label{fig:S7}
\end{figure*}

Fig.~\ref{fig:S6}A shows the resistance of device D1 as a function of temperature up to 15 K. We observe qualitatively similar behavior to device D2 (Fig.~\ref{fig:3}C of the main text), where both the resistance at $-n_s/2$ (insulator) and at slightly larger hole doping (superconductor) diverge from the high-temperature metallic response around $T$ = 5 K (Fig.~\ref{fig:S6}B). To investigate this further, we measure $dV/dI$ in device D1 as a function of $I_{dc}$ and $D$ at $-n_s/2$ (Fig.~\ref{fig:S7}A), the density where we find tunability between the insulating and superconducting phases with $D$ (see Fig.~\ref{fig:1}D of the main text). Representative cuts are shown in Fig.~\ref{fig:S7}B. Notably, the critical current of the superconductor and insulator are very similar (denoted by the red arrows), providing additional evidence of the comparable energy scales of the insulating and superconducting phases in these devices. 

\subsection*{Phase diagram in device D1}

\begin{figure*}[ht]
\includegraphics[width=6.9 in]{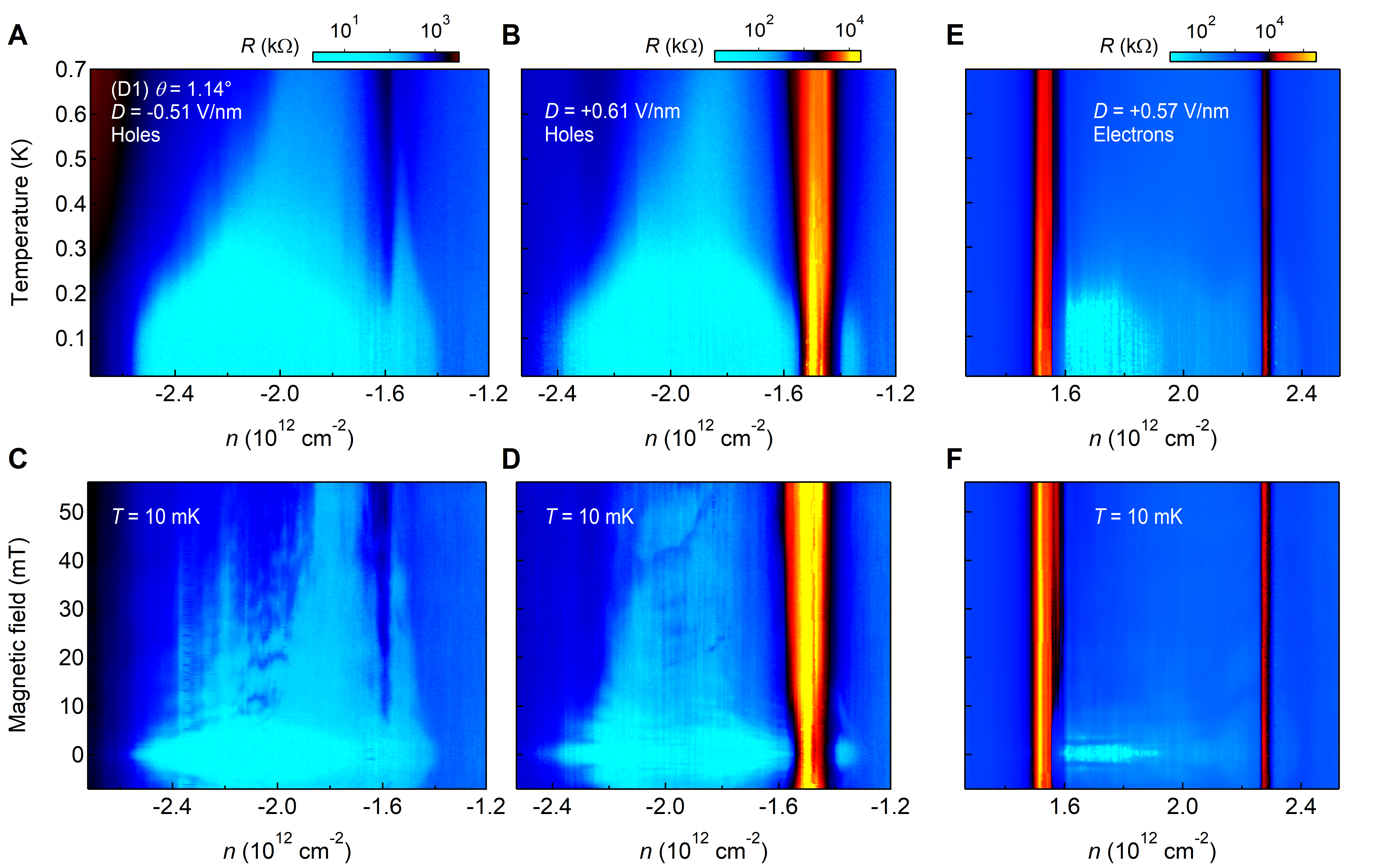} 
\caption{\textbf{Phase diagram of hole and electron carriers in device D1.}
Device resistance for hole-doping around $-n_s/2$ versus $T$ at
(\textbf{A}) $-0.51$ V/nm and 
(\textbf{B}) $+0.61$ V/nm.
We find that at $-n_s/2$ the temperature dependence of the device resistance resembles a superconductor at negative $D$ and an insulator at positive $D$.
(\textbf{C}) and (\textbf{D}) show comparable maps to (A) and (B) versus $B$. We observe oscillations in the device resistance with $B$ above the apparent $H_{c2}$.
(\textbf{E}) and (\textbf{F}) show comparable maps for electron-doping around $+n_s/2$.
}
\label{fig:S8}
\end{figure*}

Figs.~\ref{fig:S8}A-D show the resistance of device D1 for hole-doping around $-n_s/2$ as a function of temperature (panels A and B) and magnetic field (panels C and D). Figs.~\ref{fig:S8}A and C are acquired at negative $D$, while Figs.~\ref{fig:S8}B and D are acquired at positive $D$. Consistent with our observations in Fig.~\ref{fig:1}D of the main text, we find that the temperature dependence of the state at $-n_s/2$ resembles a superconductor at negative $D$ and an insulator at positive $D$. In contrast, $T_c$ appears to be independent of $D$. Figs.~\ref{fig:S8}E-F show comparable phase diagrams for electron-doping around $+n_s/2$ at positive $D$ (there is less dependence on $D$ for electron-doping). 

Figs.~\ref{fig:S8}C, D, and F exhibit oscillations in the device resistance with $B$ above the apparent $H_{c2}$, indicative of quantum interference effects arising from disorder and consistent with the observed Fraunhofer-like patterns shown in Figs.~\ref{fig:1}E-G of the main text. A small pocket of superconductivity also appears to develop for hole doping slightly less than $-n_s/2$, however this is likely a residual effect of the same disorder effects which drive the anomalous superconducting phase at $-n_s/2$ for negative $D$. 

\subsection*{Quantum oscillations in device D1}

\begin{figure*}[ht]
\includegraphics[width=3.0 in]{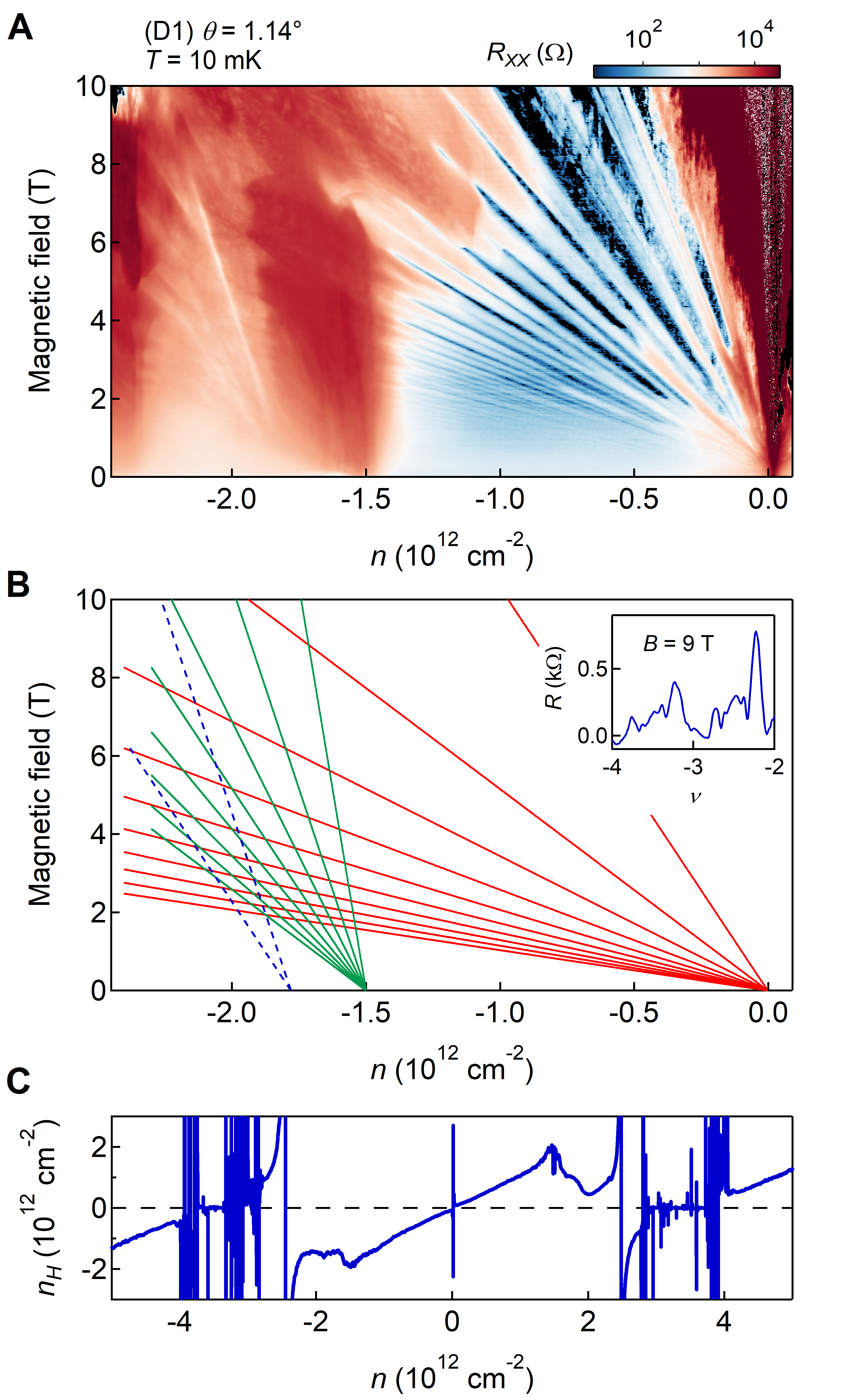} 
\caption{\textbf{Quantum oscillations in device D1.}
(\textbf{A}) Landau fan diagram of device D1 for hole-type carriers at $T$ = 10 mK. Quantum oscillations at low field emerge from the CNP with dominant degeneracy sequence of $\nu = -4, -8, -12,...$. Fractional quantum Hall states emerge for fields above $\sim$ 4 T. Separate quantum oscillations emerge from $-n_s/2$ and from roughly $-2n_s/3$, the latter of which likely corresponds to $-n^{\prime}_s/2$ of a slightly different moir\'e period within the device.
(\textbf{B}) Schematic Landau level structure from (A). Only the Landau levels persisting to the lowest fields are plotted for states emerging from the CNP, while all visible states are plotted for quantum oscillations emerging from higher filling. (inset) $R_{xx}$ between $\nu = -2$ to $-4$ averaged from $B$ = 8 to 10 T, showing the development of fractional quantum Hall states.
(\textbf{C}) Hall density $n_H$ extracted from the low-field Hall resistance as a function of total charge density induced by the gate. 
}
\label{fig:S9}
\end{figure*}

Fig.~\ref{fig:S9}A shows $R_{xx}$ as a function of $B$ up to 10 T for hole-type carriers in device D1. Fig.~\ref{fig:S9}B shows the corresponding schematic version of the observed Landau levels which persist to the lowest magnetic fields, while Fig.~\ref{fig:S9}C shows the corresponding Hall density $n_H$ as a function of total carrier density $n$. Similar to device D2, we observe a four-fold sequence of quantum oscillations emerging from the CNP with dominant filling-factor sequence $\nu = -4, -8, -12,...$ (red lines). We also observe a fully symmetry-broken sequence of quantum oscillations emerging from $-n_s/2$, shown schematically in the green lines, however these are very weak and it is difficult to determine the low-field degeneracy. We also observe two strong quantum oscillations with filling factors $\nu = -2$ and $-4$ emerging from roughly $-2n_s/3$, without a corresponding insulating state. Additionally, the quantum oscillations emerging from the CNP penetrate through those emerging from larger values of $n_s$. We believe that this unusual sequence of quantum oscillations is qualitatively consistent with our understanding of multiple moir\'e domains within this device. For example, the sequence emerging from $\sim-2n_s/3$ is likely arising from $-n^{\prime}_s/2$ of a slightly larger moir\'e unit cell somewhere within the device.

Despite the obvious signatures of structural disorder in this device, we note that the device exhibits signatures of very low charge disorder. In particular, we observe the fractional quantum Hall effect in the $N$ = 0 Landau level at fields below 4 T (Fig.~\ref{fig:S9}A). The inset of Fig.~\ref{fig:S9}B shows a cut of $R_{xx}$ at 9 T (averaged from 8 to 10 T) from $\nu = -2$ to $-4$. Fractional quantum Hall states with denominators 3 and 5 are clearly visible. 

\subsection*{Theoretical modeling of tBLG}

Since the first reports of correlated phases in tBLG emerged~\cite{Cao2018a,Cao2018b}, significant theoretical effort has focused on understanding the structure of the flat electronic bands in tBLG, as well as the nature of the insulating and superconducting phases. We briefly summarize these efforts to date here. Refs.~\cite{Zhang2018,Yuan2018,Po2018,Zhang2018b,Ray2018,Kang2018,Koshino2018,You2018,Pal2018,Guinea2018,Zou2018,Song2018,Hejazi2018,Po2018b,Tarnopolsky2018,Ahn2018} focus on the band structure of tBLG, discussing details of tight-binding models, symmetry and topology considerations of the bands, and Wannier orbital constructions. Refs.~\cite{Zhang2018,Guo2018,Yuan2018,Po2018,Padhi2018,Dodaro2018,Huang2018,Liu2018,Xu2018b,Fidrysiak2018,Rademaker2018,Isobe2018,Kennes2018,Wu2018c,Pizarro2018,Ochi2018,Thomson2018,Sherkunov2018} discuss possible mechanisms for the correlated insulating states, while Refs.~\cite{Guo2018,Xu2018,Roy2018,Baskaran2018,Huang2018,Zhang2018b,Ray2018,Liu2018,Fidrysiak2018,Rademaker2018,Isobe2018,Kennes2018,You2018,Gonzalez2018,Su2018,Sherkunov2018,Dodaro2018,Peltonen2018,Wu2018b,Ochi2018,Lian2018,Laksono2018} discuss possible mechanisms for the superconductivity. We note that there may also be some overlap in scope within these references that is not acknowledged here.

\end{document}